\documentclass[aps,prd,preprint,eqsecnum,preprintnumbers,nofootinbib,byrevtex,showpacs,showkeys,groupedaddress,amssymb,amsmath,floatfix]{revtex4}
\usepackage{bm}
\usepackage{graphics}
\usepackage{graphicx}
\usepackage{epsfig}
\usepackage{booktabs,multirow}
\usepackage{hhline}
\usepackage{slashed}
\usepackage{rccol}
\usepackage{dcolumn}
\usepackage[capitalize]{cleveref}

\begin{document}
\title{Electroweak Corrections to the  Neutralino Pair Production at CERN LHC}
\author{A.~I.~Ahmadov$^{1,2}$}
\email{E-mail:ahmadovazar@yahoo.com}
\author{M.~Demirci$^{1}$}
\email{E-mail:mehmetdemirci@ktu.edu.tr}
\affiliation{$^{1}$Department of Physics, Karadeniz Technical University, 61080 Trabzon, Turkey \\
$^{2}$ Department of Theoretical Physics, Baku State University, Z. Khalilov
St. 23, AZ-1148, Baku, Azerbaijan}%
\date{\today}
\begin{abstract}
We apply the leading and sub-leading electroweak (EW) corrections to
the Drell-Yan process of the neutralino pair production at proton-proton collision,
in order to calculate the effects of the these corrections on the neutralino pair production
at the LHC. We provide an analysis of the dependence of
the Born cross-sections for $pp\rightarrow\widetilde\chi_{i}^{0}\widetilde\chi_{j}^{0}$ and the EW corrections to this process, on the center-of-mass
energy $\sqrt s$, on the $M_2$-$\mu$ mass plane and on the squark mass for the three different scenarios. The numerical results show that the relative correction can be
reached the few tens of percent level as the increment of the center-of-mass energy, and the evaluation of EW corrections is a crucial task for all accurate measurements of the neutralino pair production processes.
\end{abstract}

\maketitle

\section{\bf Introduction}
Supersymmetry (SUSY) \cite{1971,1971_1,1971_2,1971_3,1971_4} arose as a response to attempts by
physicists to obtain a unified description of all fundamental
interaction of nature and it is at present one of the most favoured
ideas for new physics beyond the Standard Model (SM)
\cite{SusyBooks,SusyBooks2}. The realistic extension of the SM, the Minimal
Supersymmetric Standard Model (MSSM) so that it is constructed by declaring the superpartners (sparticles) of the SM states, and declaring an additional Higgs doublet (higgsinos) which has
opposite hypercharge according to Higgs doublet in the SM, so as to
give separately masses to isospin up- and down-type chiral
fermions and cancel the gauge anomalies \cite{Haber,Nilles}. The
MSSM contains a discrete symmetry known as R-parity
\cite{Ellis,Martin,Diehl,Kazakov1,Kazakov2} so that it ensures lepton and
baryon number conservations. Assuming that conservation $R$-parity,
the lightest supersymmetric particle (LSP) is definitely stable and
this particle is the end product of any process involving
sparticle in the final state. In most cases, the
stable LSP is the lightest neutralino, which is one of the
superpartners of the electroweak (EW) gauge bosons (gauginos) and the
Higgs doublet (higgsinos), which mix to form four neutral
(neutralinos $\widetilde{\chi}_{i}^{0}$) and two charged
(charginos $\widetilde{\chi}_{j}^{\pm}$) mass eigenstates.
The higgsino and gaugino decomposition of the neutralinos and
charginos includes significant information about the SUSY-breaking
mechanism and also plays an important role in the explanation of
the relic density of the dark matter \cite{DarkMatter,DarkMatter2,DarkMatter3,DarkMatter4}. Thus a
detailed study of the production of the lightest neutralino
$\widetilde{\chi}_{1}^{0}$ and the next-to-lightest neutralino
$\widetilde{\chi}_{2}^{0}$ at present and future experiments is so
important that the neutralino sector can be help us to decide
which kind of the supersymmetric models really exists in nature.

In the literature, some of the studies related to neutralino pair
production in the MSSM as follows: The neutralino pair production
via quark-antiquark annihilation at LHC was investigated in
Ref.~\cite{Ahmadov,qqbar,qqbar2}. The neutralino and chargino pair
production via gluon-gluon fusion were studied in Ref.~\cite{Jiang,Wen}
in the framework of minimal supergravity (mSUGRA) scenario. Also,
the neutralino pair production including the tree level
contributions and the leading-log one loop radiative corrections
were considered in Ref.~\cite{Gounaris}. The production of
charginos, neutralinos, and sleptons in the direct channels $p\bar{p}/pp\to
\widetilde\chi_{i}^{0}\widetilde\chi_{j}^{0}+X$ at the hadron colliders
Tevatron and LHC, via quark-antiquark annihilation was analyzed at the next-to-leading order in Ref.
\cite{Beenakker}. Focusing on the correlation of beam
polarization, the gaugino pair production in unpolarized and
polarized hadron collisions was studied in Ref.~\cite{Debove}. Moreover, the
effects of the s-channel Higgs bosons exchange on the chargino and
neutralino pair production in proton-proton collision in the
following channels $p\bar{p}/pp\to
\widetilde\chi_{i}^{0}\widetilde\chi_{j}^{0}+X$ have been analyzed
in Ref.~\cite{Arhrib}.

We analyze the dependence of the Born cross-sections and the EW corrections on the SUSY parameters for the direct production of neutralino pair at the LHC energies. One of the important approach of our scenario consist of the mechanism the choosing of input parameters. We recover the Lagrangian parameters as direct analytical expressions of suitable physical masses without any constrained in the MSSM, in such a way that we essentially focus on the algebraically
nontrivial inversion for the gaugino mass parameters, i.e., using $\tan\beta$ and two chargino masses as input parameters,
one can be obtained the other parameters, which are gaugino/higgsino mass parameters,
neutralino masses and mixing matrix as outputs. We have not only taken into account the process
$pp\rightarrow\widetilde\chi_{i}^{0}\widetilde\chi_{j}^{0}$ at the
Born level, but also logarithmic EW contributions to that process at the one-loop level.
The overall Born level magnitude of the amplitudes is reduced by these EW corrections as an amount
that could lie the few tens of percent level for the kinematical
domain attainable at the LHC. Therefore, these corrections are important
for the experimental and theoretical studies related to the
production of neutralino pair at the LHC
and the future colliders.

The remainder of this paper is organized as follows: In Section
\ref{parameters}, we present briefly definitions corresponding to
the neutralino/chargino sector and our method for calculations.
In Section \ref{cs}, the analytical expressions of the amplitudes
and the cross-sections is given for subprocess
$q\bar{q}\to \widetilde\chi_{i}^{0}\widetilde\chi_{j}^{0}$. In
Section \ref{corrections}, we provide the formulas of the leading
and subleading EW logarithmic corrections for amplitudes of the
subprocess $q\bar{q}\to
\widetilde\chi_{i}^{0}\widetilde\chi_{j}^{0}$ and in Section
\ref{results}, the numerical results for the
cross-section and the EW corrections is given, and we discuss the dependence of the cross-section on the SUSY
model parameters. Finally, our conclusions are given in section \ref{Conc}.

\section{The neutralino/chargino sector of the MSSM}\label{parameters}

The  physical neutralino mass eigenstates
$\widetilde\chi_i^{0}$ ($i=1,..,4$) are the combinations of the
neutral gauginos $\widetilde B$, $\widetilde{W}^{3}$ and the
neutral higgsinos $\widetilde H_1^{0}$, $\widetilde H_2^{0}$ in the MSSM. The
soft SUSY-breaking terms in the Lagrangian include the following term
\cite{Haber},
\begin{equation} \label{eq:Lagrangian}
\mathcal{L} ~\supset~ -\frac{1}{2}(\psi^0_i)^{T} \mathcal{M}
\psi^0_j +h.c.,
\end{equation}
which is bilinear in the fermion fields
${\psi}_{j}^{0}=(-i \widetilde B,-i \widetilde W^3,\widetilde
H_1^{0},\widetilde H_2^{0})^T$ with $j=1,2,3,4$. In the above
relation, the neutralino mass matrix is given as
\begin{equation} \label{eq:MassMatrixN}
\mathcal{M}=\left(\begin{array}{cccc}M_{1}&0&-m_{Z}c_{\beta}s_{W}&m_{Z}s_{\beta}s_{W}\\
0&M_{2}&m_{Z}c_{\beta}c_{W}&-m_{Z}s_{\beta}c_{W}\\
-m_{Z}c_{\beta}s_{W}&m_{Z}c_{\beta}c_{W}&0&-\mu\\
m_{Z}s_{\beta}s_{W}&-m_{Z}s_{\beta}c_{W}&-\mu&0\end{array}\right),
\end{equation}
which is symmetric. Here, $\mu$ and $M_{1}$/$M_{2}$ are the
supersymmetric Higgssino mass parameter and the gaugino
mass parameter related to the $U(1)$/$SU(2)$ subgroup, respectively, and $tan\beta
=v_2/v_1$ is the ratio of the vacuum expectation values of the two Higgs fields that
break the EW symmetry. The mass parameters are possibly complex
in CP noninvariant theories, in this case, by means of the reparametrization
of the fields, the $M_{2}$ gaugino mass can be obtained as real and
positive with no loss of generality in order that the two remaining
nontrivial phases, which are reparametrization invariant, can be
ascribed to $\mu$ and $M_{1}$ as follows: $\mu=|\mu|e^{i\phi_{\mu}}$
and $ M_{1} =|M_{1}|e^{i\phi_{1}}$ $(\phi_\mu<2\pi, 0\leq\phi_1)$.

The neutralino mass matrix $\mathcal{M}$ can be diagonalized by one
$4\times4$ unitary matrix $N$, which is sufficient to rotate from
the gauge eigenstate basis ($\widetilde{B}^0,\widetilde{W}^3,\widetilde {H}_{1}^0,\widetilde
{H}_{2}^0$) to the mass eigenstate basis of the neutralino fields
$\widetilde{\chi}_{i}^{0}$:
\begin{equation} \label{eq:Md}
\mathcal{M}_D=N^{T}\mathcal{M}N=\sum_{j=1}^{4}m_{\widetilde{\chi}_{j}^{0}}E_{j}.
\end{equation}
Therefore, the relation between physical and weak eigenstates can be
extracted as $\chi_{i}^{0}=N_{ij}{\psi}_{j}^{0}$ with $i=1,2,3,4$.
In order to determine $N$, the square of Eq.~\eqref{eq:Md} obtaining
\begin{equation} \label{eq:Md2}
\mathcal{M}_{D}^2=N^{-1}\mathcal{M}^{+}\mathcal{M}N=\sum_{j=1}^{4}m_{\widetilde{\chi}_{j}^{0}}^2E_{j},
\end{equation}
where $(E_j)_{ik}=\delta_{ji}\delta_{jk}$. The neutralino mass eigenstates are expressed by
\begin{equation}
\widetilde{\chi}_{j}^{0}=\left(\begin{array}{cc} \chi_{j}^{0}\\
\overline{\chi}_{j}^{0}\\ \end{array}\right),
\end{equation}
where $\chi_{j}^{0}$ denotes the two component Weyl spinor and
$\widetilde{\chi}_{j}^{0}$ the four component Majorana spinor of the
$j$th neutralino field. The application of projection operators
leads to relatively compact analytic expressions for the mass
eigenvalues $ m_{\widetilde{\chi}_{1}^{0}}<
m_{\widetilde{\chi}_{2}^{0}} < m_{\widetilde{\chi}_{3}^{0}}<
m_{\widetilde{\chi}_{4}^{0}}$ \cite{Gounaris2}. The mass eigenvalues
$m_{\widetilde{\chi}_{j}^{0}}$ in the diagonal neutralino mass matrix $\mathcal{M}_{D}$ are possibly chosen as
positive and reel by an appropriate definition of the unitary matrix
$N$. Rearranging Eq.~\eqref{eq:Md2} as follows,
\begin{equation}
(\mathcal{M}^{+}\mathcal{M})N-N\mathcal{M}_{D}^2=0
\end{equation}
 and by solving this system of equations and by taking into account the
 following relation
\begin{equation}
|N_{1j}|^2+|N_{2j}|^2+|N_{3j}|^2+|N_{4j}|^2=1,
\end{equation}
the $N_{ij}$ matrix's components are obtained. Also, the neutralino
masses are obtained by solving the following characteristic equation,
\begin{equation}\label{eq:charac}
X^4-aX^3+bX^2-cX+d=0,
\end{equation}
where
$$
a =  M_1^2+2\mu^2+M_2^2+2m_Z^2,
$$
$$
b =
(\mu^2+m_Z^2)^2+M_2^2(M_1^2+2\mu^2+2m_{Z}^2s_W^2)+2M_1^2(\mu^2+m_Z^2c_W^2)-2\mu
m_Z^2c_W^2 M_2sin2\beta
$$
$$
\times cos\phi_{\mu}-2m_Z^2s_W^2M_1 sin2\beta
cos(\phi_{\mu}+\phi_1),
$$
$$
c =  {\mu}^4 M_{1}^2+{\mu}^2 m_{Z}^4 sin^2
2\beta+M_1^2m_Z^2c_W^2(2\mu^2+m_Z^2c_W^2)+
$$
$$
M_2^2(m_{Z}^4s_{W}^4+2\mu^2(m_Z^2s_W^2+M_1^2)+\mu^4)-2\mu
m_Z^2s_W^2M_1(\mu^2+M_2^2)sin2\beta cos(\phi_{\mu}+\phi_1)+
$$
$$
2m_Z^2 c_W^2 M_2[m_Z^2M_1s_W^2cos\phi_1-\mu(\mu^2+M_1^2)
cos\phi_{\mu} sin2\beta],
$$
$$
d =  m_Z^4 c_W^4\mu^2 M_1^2
sin^2{2\beta}+2m_Z^2\mu^2M_1M_2c_W^2(m_Z^2s_W^2sin2\beta
cos\phi_1-\mu M_1cos\phi_\mu) +
$$
$$
\mu^2m_Z^2s_W^2M_2^2sin2\beta(m_Z^2s_W^2sin2\beta-2\mu
M_1cos(\phi_1+\phi_{\mu}))+\mu^4 M_1^2M_2^2.
$$
From solving Eq.~\eqref{eq:charac}, the exact analytic formulas of
the neutralino masses are obtained as follows,
\begin{equation} \label{eq:mN}
\begin{split}
  & m_{\widetilde{\chi}_{1}^{0}}^2,
m_{\widetilde{\chi}_{2}^{0}}^2=\frac{a}{4}-\frac{f}{2}\mp\frac{1}{2}\sqrt{r-w-\frac{p}{4f}}, \\
  & m_{\widetilde{\chi}_{3}^{0}}^2,
m_{\widetilde{\chi}_{4}^{0}}^2=\frac{a}{4}+\frac{f}{2}\mp\frac{1}{2}\sqrt{r-w+\frac{p}{4f}}.
\end{split}
\end{equation}
where
\begin{equation}
\begin{split}
  & f=\sqrt{\frac
{r}{2}+w},~~r=\frac{a^2}{2}-\frac{4b}{3},~~w=\frac{q}{(3\cdot
2^{1/3})}+\frac{(2^{1/3}\cdot h)}{3\cdot q} \\
  & p=a^3-4ab+8c,~~q=(k+{\sqrt{k^2-4h^3}})^{1/3} \\
  & k=2b^3-9abc+27c^2+27a^2d-72bd,~~h=b^2-3ac+12d.
\end{split}
\end{equation}

The  physical chargino mass eigenstates $\widetilde\chi_i^{\pm}$
(i=1,2) are the combinations of the charged gauginos ( $\widetilde
W^{\pm}$) and the charged higgsinos ($H_{2,1}^{\pm}$). In terms of
two-component Weyl spinors, the chargino mass term in the SUSY
Lagrangian can be expressed by \cite{Haber}
\begin{equation} \label{eq:LagrangianX}
\mathcal{L} ~\supset~ -\frac{1}{2}\left(\begin{array}{cc} \psi^+
&\psi^- \end{array}\right) \left(\begin{array}{cc} 0
&\mathcal{M}_C^T\\ \mathcal{M}_C
&0\end{array}\right) \left(\begin{array}{c} \psi^+\\
\psi^-\end{array}\right)+h.c.,
\end{equation}
which is bilinear in the two-component fermionic fields
${\psi}_{j}^{\pm}=(-i \widetilde W^{\pm},\widetilde
H_{2,1}^{\pm})^T$ with $j=1,2$.
 The chargino mass matrix $\mathcal{M}_C $ is given as
\begin{equation} \label{eq:MassMatrixC}
\mathcal{M}_C=\left(\begin{array}{cc}M_2&{\sqrt{2}}m_Wc_\beta \\
\sqrt{2}m_Ws_\beta &|\mu|e^{i\phi_\mu}\end{array}\right).
\end{equation}
The matrix $\mathcal{M}_C$ is not symmetric, so it must be
diagonalized by two different unitary matrices $V$ and $U$, which
lead to the relation
$U^*\mathcal{M}_CV^{-1}=\text{diag}\left\{m_{\widetilde{\chi}_{1}^{\pm}},m_{\widetilde{\chi}_{2}^{\pm}}\right\}$,
with the chargino mass eigenvalues:
\begin{equation} \label{eq:m_chargino}
\begin{split}
m^2_{\widetilde{\chi}_{1,2}^{+}}=&\frac{1}{2}\bigl\{M_{2}^2+{|\mu|}^2+2m^2_W\mp
\bigl[(M_2^2-{|\mu|}^2-2m^2_{W}
\cos2\beta)^2 \\
  & +8m^2_W(M_2^{2}c^2_\beta+{|\mu|}^2 s^2_\beta +
M_2|\mu|\sin2\beta \cos\phi_\mu)\bigr]^{1/2}\bigr\}.
\end{split}
\end{equation}

The fundamental SUSY parameters $M_2$ and $\mu$
are possibly derived from these two chargino masses for given $tan\beta$ \cite{Choi1,Moultaka}. By taking appropriate sum and differences of the chargino masses in the
Eq.~\eqref{eq:m_chargino}, one can be derived the following equations for
$M_2$ and $\mu$:
\begin{equation} \label{eq:2M2}
2M_2^2=(m_{\widetilde{\chi}_{1}^{+}}^2+m_{\widetilde{\chi}_{2}^{+}}^2-
2m_{W}^2)\mp\sqrt{(m_{\widetilde{\chi}_{1}^{+}}^2+m_{\widetilde{\chi}_{2}^{+}}^2-
2m_{W}^2)^2-\Delta_{\pm} },
\end{equation}
\begin{equation} \label{eq:2mu2}
2|\mu|^2=(m_{\widetilde{\chi}_{1}^{+}}^2+m_{\widetilde{\chi}_{2}^{+}}^2-
2m_{W}^2)\pm
\sqrt{(m_{\widetilde{\chi}_{1}^{+}}^2+m_{\widetilde{\chi}_{2}^{+}}^2-2m_{W}^2)^2-
\Delta_{\pm}}
\end{equation}
with
$$
\Delta_{\pm}=4 \left[m_{\widetilde{\chi}_{1}^{+}}^2
m_{\widetilde{\chi}_{2}^{+}}^2+ m_{W}^4
cos2\phi_{\mu}sin^2{2{\beta}}\pm 2m_{W}^2 cos\phi_{\mu}
sin2\beta\times \right.
$$
$$
\left.\sqrt{m_{\widetilde{\chi}_{1}^{+}}^2
m_{\widetilde{\chi}_{2}^{+}}^2-m_{W}^4 sin^2{2{\beta}}
sin^2{\phi_{\mu}}}\right].
$$
In the above equations, the upper (lower) signs correspond to
$M_2<|\mu|$ ($M_2>|\mu|$) regime. Here, four solutions associated with different physical scenarios are occurred.
For the $M_{2}>|\mu|$ regime, the lightest chargino has a stronger
higgsino-like component and thus it is mentioned as higgsino-like \cite{Moultaka,Choi}.
The solution for the $|\mu|>M_{2}$ regime corresponds to the
gaugino-like case, could be easily figured out by the following
replacements: $M_{2} \to |\mu|$ and $\mu\to$ sign($\mu$)$M_{2}$
\cite{Choi,Kneur}. The universality of the gaugino
masses at the GUT scale, which leads to the relation,
\begin{equation}
M_1=\frac{5}{3}M_2 \tan^2\theta_W.
\end{equation}

In this work, we take into account the gaugino/higgsino sector with
the following assumptions: First, in order to obtain reel mass
eigenvalues, namely $\phi_1=0$ and $\phi_{\mu}=0$. The signs
among the mass parameters $M_1$, $M_2$ and $\mu$ are relative, which can be absorbed
into phases  $\phi_1$ and $\phi_{\mu}$ by redefinition of fields,
and consequently, these mass parameters can be real and
positive. Under the these assumptions, it is possible that there
appear several scenarios for the choice of the parameters. On
account of the fact that the SUSY parameters can be derived from
the physical quantities, it is also possible that choose an
alternative way to diagonalize the mass matrix, by
using two chargino masses together with $\tan\beta$ as inputs.
Moreover, there are several scenarios for the choice of two
chargino masses and $tan\beta$ \cite{Kneur}. The scenarios
correspond to the choice of $tan\beta$ as follow: scenario with
small $tan\beta$ ($tan\beta\approx1\div3$) and scenario with large
$tan\beta$ ($tan\beta\approx30\div70$) \cite{Barger,Langacker,Kelley,Gladyshev}.

\section{Calculation of the cross section }\label{cs}
In this section, we present analytical expressions of amplitudes and
the cross-section of the neutralino pair production. The neutralino
pair production originates from quark-antiquark collision, is expressed by
\begin{equation} \label{eq:qqnn}
q(p_1)\overline q(p_2)\rightarrow\widetilde\chi_{i}^{0}(k_1)\widetilde\chi_{j}^{0}(k_2),
\end{equation}
where $p_{1}$, $p_{2}$, $k_1$ and $k_2$ represent the four momenta
of the quark, antiquark, the two final state neutralinos, separately. The
Mandelstam variables for subprocess are given by
\begin{equation}
\hat s=(p_1+p_2)^2, \quad \hat t=(p_1-k_1)^2,\quad \hat
u=(p_1-k_2)^2.
\end{equation}
The relevant couplings of the supersymmetric particles for
neutralino pair production are extracted from the following
interaction Lagrangians \cite{Rosiek} so that,
\begin {equation} \label{eq:LZNN}
L_{Z^{0}\widetilde{\chi}_{i}^{0}\widetilde{\chi}_{j}^{0}}=
\frac{1}{2}\frac{g}{cos\theta_{\rm W}}
Z_{\mu}\overline{\widetilde{\chi}}_{i}^{0}\gamma^{\mu}\left(O_{ij}^{\prime\prime}
 P_{L}+O_{ij}^{\prime\prime}
P_{R}\right)\widetilde{\chi}_{j}^{0},
\end {equation}
\begin {equation} \label{eq:LZqq}
L_{Z^{0}q\bar{q}}=\frac{g}{cos\theta_{\rm W}} \bar
{q}\gamma^{\mu}\left( L_{q} P_{L}+R_{q} P_{R}\right)q Z_{\mu},
\end {equation}
\begin {equation} \label{eq:LqsqN}
L_{q\widetilde{q}\widetilde{\chi}^{0}}=\bar{q}\left(a_{i}^{L}(\widetilde{q}_{n})P_L
+a_{i}^{R}(\widetilde{q}_{n})P_R\right)\widetilde{\chi}_{i}^{0}\widetilde{q}_{n},
\end {equation}
where $q$, $\widetilde{q}_n$ and $\widetilde{\chi}_{i}^{0}$ denote four-component spinor
fields of the quark, squark and neutralino, respectively. Moreover,
$g=e/sin\theta_{W}$ is the weak coupling constant, $P_{R,L}=\frac{1}{2}(1\pm\gamma^5)$. In the above
Lagrangians, the relevant couplings $O_{ij}^{\prime\prime},
L_{q},R_{q}$ and $a_{i}^{R,L}(\widetilde{q}_{n})$ are given by
\begin{equation} \label{eq:o_ij}
O_{ij}^{{\prime\prime}
L}=O_{Z}^{ij}=\frac{1}{2}(N_{i3}N_{j3}^{\star}-N_{i4}N_{j4}^{\star})cos2\beta
-\frac{1}{2}(N_{i3}N_{j4}^{\star}+N_{i4}N_{j3}^{\star})sin2\beta,
\end{equation}
\begin{equation}
O_{ij}^{{\prime\prime} R}=-O_{Z}^{ij\star},
\end{equation}
\begin{equation}
L_{q}=2I_{q}^3(1-2sin^2\theta_{W}|Q_q|),\,\,\,R_{q}=-2sin^2\theta_{W}Q_q,
\end{equation}
with $I_{q}^3,Q_q$ which are the isospin quantum number and charge
of the various quarks, and
\begin{equation} \label{eq:a_i}
\begin{split}
& a_i^L(\widetilde{u}_L)=-\frac{e}{{3\sqrt 2}s_W
c_W}(N_{1i}s_W+3N_{2i}c_W),~~ a_i^L(\widetilde{u}_R)=-\frac{e
m_u}{{\sqrt 2} m_W s_W s_{\beta}}N_{4i},\\
  & a_i^R(\widetilde{u}_R)=\frac{2{\sqrt 2}e}{3 c_W}N_{1i}^{\star},~~
a_i^R(\widetilde{u}_L)=-\frac{e m_u}{{\sqrt 2} m_W s_W
s_{\beta}}N_{4i}^{\star},\\
  &a_i^L(\widetilde{d}_L)=-\frac{e}{{3\sqrt 2}s_W c_W}(N_{1i}
s_W-3N_{2i} c_W),~~ a_i^L(\widetilde{d}_R)=-\frac{e m_d}{{\sqrt 2}
m_W s_W c_{\beta}}N_{3i},\\
  & a_i^R(\widetilde{d}_R)=-\frac{{\sqrt 2}e}{3 c_W} N_{1i}^{\star},~~ a_i^R(\widetilde{d}_L)=-\frac{e m_d}{{\sqrt 2} m_W
s_W c_{\beta}}N_{3i}^{\star}.
\end{split}
\end{equation}
One can note that the mixing matrices $N_{ij}$ control the higgsino and gaugino
components of the neutralino in the $Z\widetilde{\chi}_{i}^{0}\widetilde{\chi}_{j}^{0}$ and
$q\widetilde{q}\widetilde{\chi}^{0}$ coupling as shown in the Lagrangians.
\begin{figure}[ht]
     \begin{center}
\epsfig{figure=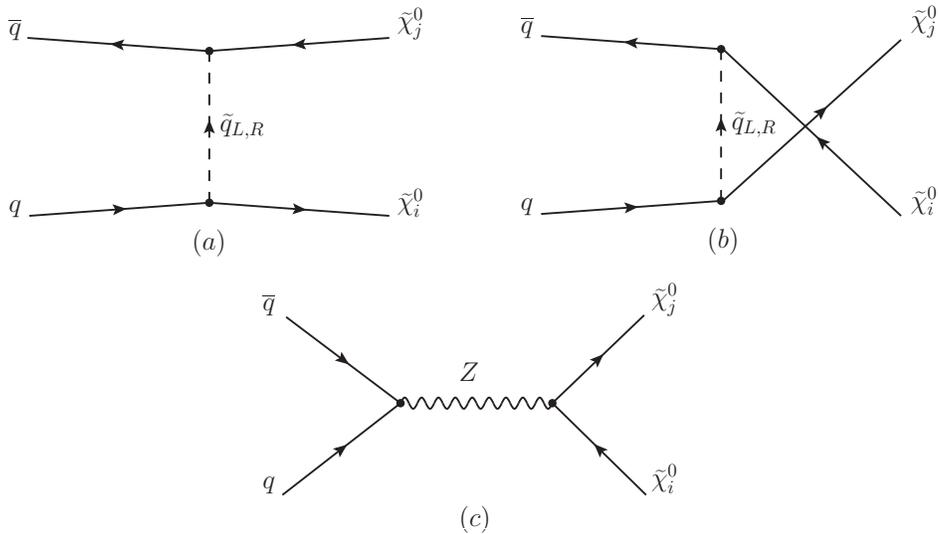,width=0.8\textwidth}
     \end{center}
\caption{Feynman diagrams of the subprocess $q\bar{q}
\to\widetilde{\chi}_{i}^{0}\widetilde{\chi}_{j}^{0}$ to leading level.} \label{Fig1}
\end{figure}

The subprocess for neutralino pair production proceeds through \textit{t}- and \textit{u}-channel contributions due to exchange of the
squarks, and \textit{s}-channel contribution due to $Z$ boson exchange as shown in~\cref{Fig1}. The corresponding amplitudes for each diagram can be given as
\begin{equation} \label{eq:matrixelement}
T=T_{\hat s} +T_{\hat t} +T_{\hat u},
\end{equation}
where
\begin{equation} \label{eq:ampl}
\begin{split}
  T_{\hat s}=&-\frac{e^2}{2sin^2\theta_W cos^2\theta_W}D_{Z}(\hat s)\overline{u}_{i}(k_1)
\gamma^{\mu}\left[O_{Z}^{ij}P_L -
O_{Z}^{ij\star}P_R\right]{\vartheta}_{j} (k_2)\\
  &\times\overline{v}(p_2)\gamma_{\mu}\left[g_{V_{q}}+g_{A_{q}}\gamma_5\right]u(p_1),\\
T_{\hat t}= &\sum_{n}\frac{1}{\hat t-m_{\widetilde
q_n}^2}\bar{u}_{i}(k_1)\left[a_{i}^{L}(\widetilde{q}_{n})P_L+a_{i}^{R}(\widetilde{q}_{n})P_R\right]u(p_1)\\
  & \times\bar{v}(p_2)\left[a_{j}^{L\star}(\widetilde{q}_{n})P_R
+a_{j}^{R\star}(\widetilde{q}_{n})P_L\right]v_{j}(k_2),\\
T_{\hat u}=&-\sum_{n}\frac{1}{\hat u - m_{\widetilde q_n}^2}
\bar{u}_{j}(k_2)\left[a_{j}^{L}(\widetilde{q}_{n})P_L +
a_{j}^{R}(\widetilde{q}_{n})P_R\right]u(p_1)\\
 &\times\bar{v}(p_2)\left[a_{i}^{L\star}(\widetilde{q}_{n})P_R +
a_{i}^{R\star}(\widetilde{q}_{n})P_L\right] v_{i}(k_1),
\end{split}
\end{equation}
where the label $n$ denotes the summation over the exchanged
$\widetilde{q}_L$ and $\widetilde{q}_R$ squarks of the same flavor
in the $t$-and $u$- channel, and $i,j$ denote the type of the final state
neutralinos. After averaging over spins and colors in the initial
state, the unpolarized differential cross-section is given by
\begin{equation}
\frac{d\hat \sigma (q\overline
q\rightarrow\widetilde\chi_{i}^{0}\widetilde\chi_{j}^{0})}{d\hat
t}=\frac{1}{16\pi \hat
s^2}\frac{1}{3}\frac{1}{4}\left(\frac{1}{2}\right)^{\delta_{ij}}\left(M_{\hat
s \hat s}+M_{\hat t \hat t}+M_{\hat u \hat u} -2M_{\hat s \hat t}
+2M_{\hat s \hat u}-2M_{\hat t \hat u}\right),
\end{equation}
where the factors $\frac{1}{3}$, $\frac{1}{4}$ and
$(\frac{1}{2})^{\delta_{ij}}$ come from averaging over color, spin
in the initial state and the final identical particle factor,
respectively. The squares of the amplitudes can be obtained and
summed over final states using standard trace techniques. Therefore,
we obtain the following equations,
\begin{equation}
\begin{split}
M_{\hat s \hat s} =& \frac{e^4}{4\sin^4\theta_W
\cos^4\theta_{W}}|D_{Z}(\hat s)|^2 (L_{q}^2+R_{q}^2)\biggl\{
{O_{Z}^{ij}O_{Z}^{ij \star}}[(m_{\widetilde{\chi}_{i}^{0}}^2 - \hat
u)(m_{\widetilde{\chi}_{j}^{0}}^2 -\hat u) \\
  & +(m_{\widetilde{\chi}_{i}^{0}}^2 - \hat
t)(m_{\widetilde{\chi}_{j}^{0}}^2 -\hat
t)]-m_{\widetilde{\chi}_{i}^{0}}m_{\widetilde{\chi}_{j}^{0}}\hat s
(O_{Z}^{ij2}+O_{Z}^{ij \star 2})\biggr\},
\end{split}
\end{equation}

\begin{equation}
\begin{split}
M_{\hat t \hat t} =& \sum_{k,l}\frac{1}{(\hat t - m_{\widetilde
{q}_{k}}^2)(\hat
t-m_{\widetilde{q}_{l}}^2)}\biggl\{[a_{i}^{L}(\widetilde{q}_{k})a_{i}^{L\star}(\widetilde{q}_{l})+a_{i}^{R}(\widetilde{q}_{k})a_{i}^{R\star}(\widetilde{q}_{l})]
[a_{j}^L(\widetilde{q}_{k})a_{j}^{L\star}(\widetilde{q}_{l}) \\
  & +a_{j}^R(\widetilde{q}_{k})a_{j}^{R\star}(\widetilde{q}_{l})]\biggr\}
(m_{\widetilde{\chi}_{i}^{0}}^2- \hat
t)(m_{\widetilde{\chi}_{j}^{0}}^2 - \hat t),
\end{split}
\end{equation}

\begin{equation}
\begin{split}
M_{\hat u \hat u} =& \sum_{k,l}\frac{1}{(\hat u -
m_{\widetilde{q}_{k}}^2)(\hat
u-m_{\widetilde{q}_{l}}^2)}\biggl\{[a_{i}^{L\star}(\widetilde{q}_{k})a_{i}^L(\widetilde{q}_{l})+a_{i}^{R\star}(\widetilde{q}_{k})a_{i}^R(\widetilde{q}_{l})]
[a_{j}^L(\widetilde{q}_{l}) a_{j}^{L\star}(\widetilde{q}_{k}) \\
  &+ a_{j}^R(\widetilde{q}_{l})
a_{j}^{R\star}(\widetilde{q}_{k})]\biggr\}
(m_{\widetilde{\chi}_{i}^{0}}^2- \hat
u)(m_{\widetilde{\chi}_{j}^{0}}^2 - \hat u),
\end{split}
\end{equation}

$$
M_{\hat t \hat u}=\sum_{k,l}\frac{1}{(\hat t -
{m}_{\widetilde{q}_{k}}^2)(\hat
u-{m}_{\widetilde{q_l}}^2)}\biggl\{\frac{1}{2}\left
[a_{i}^{L\star}(\widetilde{q}_{k})a_{j}^L(\widetilde{q}_{l})
a_{j}^R(\widetilde{q}_{k})
a_{i}^{R\star}(\widetilde{q}_{l})+a_{i}^{R\star}(\widetilde{q}_{k})
a_{j}^R(\widetilde{q}_{l}) \right.
$$
$$
\left.a_{i}^{L\star}(\widetilde{q}_{l})
a_{j}^L(\widetilde{q}_{k})\right]
[(m_{\widetilde\chi_{j}^{0}}^2-\hat
u)(m_{\widetilde{\chi}_{i}^{0}}^2-\hat u)+
(m_{\widetilde{\chi}_{j}^{0}}^2-\hat
t)(m_{\widetilde{\chi}_{i}^{0}}^2-\hat t)-\hat s(\hat s
-m_{\widetilde{\chi}_{i}^{0}}^2-m_{\widetilde{\chi}_{j}^{0}}^2)] +
$$
\begin{equation}
m_{\widetilde{\chi}_{i}^{0}} m_{\widetilde{\chi}_{j}^{0}}\hat
s[a_{j}^{L\star}(\widetilde{q}_{l}) a_{i}^{L}(\widetilde{q}_{k})
a_{i}^{L}(\widetilde{q}_{l}) a_{j}^{L\star}(\widetilde{q}_{k})+
a_{j}^{R\star}(\widetilde{q}_{l}) a_{i}^{R}(\widetilde{q}_{k})
a_{i}^{R}(\widetilde{q}_{l})
a_{j}^{R\star}(\widetilde{q}_{k})]\biggr\}
\end{equation}

$$
M_{\hat s \hat
u}=\sum_{k}\frac{e^2}{2sin^2\theta_Wcos^2\theta_W(\hat
u-m_{\widetilde {q}_{k}}^2)}(Re[D_{Z}(\hat s)])\biggl\{[L_q
a_{i}^{L\star}(\widetilde{q}_{k})a_{j}^{L}(\widetilde{q}_{k})
O_{Z}^{ij\star}- \\
$$
$$
R_q a_{i}^{R\star} (\widetilde{q}_{k}) a_{j}^{R}(\widetilde{q}_{k})
O_{Z}^{ij}] (m_{\widetilde{\chi}_{i}^{0}}^2-\hat
u)(m_{\widetilde{\chi}_{j}^{0}}^2 -\hat u)+ [R_q
a_{i}^{R\star}(\widetilde{q}_{k}) a_{j}^{R}(\widetilde{q}_{k})
O_{Z}^{ij\star}- \\
$$
\begin{equation}
L_q a_{i}^{L\star}(\widetilde{q}_{k}) a_{j}^{L}(\widetilde{q}_{k})
O_{Z}^{ij}] m_{\widetilde{\chi}_{i}^{0}}
m_{\widetilde{\chi}_{j}^{0}} \hat s \biggr\},
\end{equation}

$$
M_{\hat s \hat
t}=\sum_{k}\frac{e^2}{2sin^2\theta_Wcos^2\theta_W(\hat
t-m_{\widetilde q_k}^2)}(Re[D_{Z}(\hat s)])\biggl\{[R_q
a_{j}^{R\star}(\widetilde{q}_{k})a_{i}^{R}(\widetilde{q}_{k})
O_{Z}^{ij\star}- \\
$$
$$
L_q a_{j}^{L\star}(\widetilde{q}_{k}) a_{i}^{L}(\widetilde{q}_{k})
O_{Z}^{ij}] (m_{\widetilde{\chi}_{i}^{0}}^2-\hat
t)(m_{\widetilde{\chi}_{j}^{0}}^2-\hat t)+ [L_q
a_{j}^{L\star}(\widetilde{q}_{k})
a_{i}^{L}(\widetilde{q}_{k}) O_{Z}^{ij\star}- \\
$$
\begin{equation}
R_q a_{j}^{R\star}(\widetilde{q}_{k}) a_{i}^{R}(\widetilde{q}_{k})
O_{Z}^{ij}]m_{\widetilde{\chi}_{i}^{0}}
m_{\widetilde{\chi}_{j}^{0}} \hat{s}\biggr\},
\end{equation}
In the above equations, the following abbreviation is used
\begin{equation}
D_{Z}(\hat s)=\frac{1}{\hat s - m_{Z}^2+im_{Z} \Gamma_{Z}}
\end{equation}
for propagator of the boson $Z^0$. We get $m_{Z^0}$ = 91.1876 GeV
and the width of the boson $Z^0$ by $\Gamma_{Z}=2.499947 $ GeV. To
obtain the final cross-section, we use the basic parton model
expression of the hadron-hadron collision $h_1(p_1)h_2(p_2)\to
\widetilde{\chi}_{i}^{0}(k_i)\widetilde{\chi}_{j}^{0}(k_j)$
\cite{Owens,Greiner} which is
\begin{equation} \label{eq:dsigma}
\frac{d\sigma}{d \cos\theta}=\frac{1}{2} \sum_{q_{1}q_{2}}\int\int
dx_{1} dx_{2}~x_{1}G_{{q_{1}}/{h_{1}}}(x_{1},Q)
~x_{2}G_{{q_{2}}/{h_{2}}}(x_{2},Q)\frac{d\hat \sigma (q_{1}
q_{2}\rightarrow\widetilde\chi_{i}^{0}\widetilde\chi_{j}^{0})}{d\hat
t},
\end{equation}
where $G_{q_1/h_{1}}(x_1,Q)$ ($G_{q_2/h_{2}}(x_2,Q)$) is the
distribution function of parton $q_1$ ($q_2$) in the hadron $h_1$ ($h_2$)
 at the factorization scale Q. We fix the factorization scale to
the average mass of the final state particles, $Q=(m_{\widetilde{\chi}_{i}^{0}}+m_{\widetilde{\chi}_{j}^{0}})/2$. Taking the
$h_{1}h_{2}$-center-of-mass system as the Lab-system, the
Lab-momentums of the produced $\widetilde{\chi}_{i}^{0}$ and
$\widetilde{\chi}_{j}^{0}$ are \cite{Bycling}
\begin{equation}
k_{i}^{\mu}=(E_i,k_{T}, k_{i}cos\theta),\,\,\,
k_{j}^{\mu}=(E_j,-k_T, k_{j} cos\theta),
\end{equation}
where their transverse momentums are clearly just opposite such that
$k_T=k_{T_i}=-k_{T_j}$, while their transverse energies
$E_{T_i}=\sqrt{k_{T}^2+m_{\widetilde{\chi}_{i}^{0}}^2}$,
$E_{T_j}=\sqrt{k_T^2+m_{\widetilde{\chi}_{j}^{0}}^2}$ are used to
define $x_{T_{i,j}}=2E_{T_{i,j}}/\sqrt{s}$. Moreover, the momentums
of the incoming partons are expressed by
\begin{equation}
p_1=\frac{\sqrt{s}}{2}(x_1,0,0,x_1),\,\,\,p_2=\frac{\sqrt{s}}{2}(x_2,0,0,-x_2),\,\
\end{equation}
\begin{equation}
p^0=\frac{\sqrt s}{2}(x_1+x_2)=E_i+E_j,\,\,\,p_3=\frac{\sqrt
s}{2}(x_1-x_2)=(k_icos\theta_i+k_jcos\theta_j),
\end{equation}
which lead to
\begin{equation}
x_1=\frac{1}{2}[x_{T_i} e^{y_i}+x_{T_j} e^{y_j}]=\frac{M}{\sqrt
s}e^{\bar y},
\end{equation}
\begin{equation}
x_2=\frac{1}{2}[x_{T_i} e^{-y_i}+x_{T_j} e^{-y_j}]=\frac{M}{\sqrt
s}e^{-\bar y},
\end{equation}
\begin{equation}
\hat s=M^2=(p_1+p_2)^2=x_{1}
x_{2}s=\frac{s}{4}[x_{T_i}^2+x_{T_j}^2+2x_{T_i}x_{T_j}cosh({\Delta
y})].
\end{equation}

Using Eq.~\eqref{eq:dsigma}, the  expression the differential cross-section
in terms of the overall center-of-mass rapidities of the two
jets is obtained as follows,
\begin{equation}
\frac{d\sigma}{dy_{i}dy_{j}dk_{T}^2}=x_{1}x_{2}\sum_{q_1
q_2}G_{{q_{1}}/{h_{1}}}(x_{1},Q)
 G_{{q_{2}}/{h_{2}}}(x_{2},Q)\frac{d\hat \sigma (q_{1}
q_{2}\rightarrow\widetilde\chi_{i}^{0}\widetilde\chi_{j}^{0})}{d\hat
t}.
\end{equation}
\section{Electroweak logarithmic corrections on the amplitudes of the subprocesses $q\bar{q}\to
\widetilde\chi_{i}^{0}\widetilde\chi_{j}^{0}$ at
one-loop}\label{corrections}

In the TeV range such terms reach the several percent level and be
easily measurable at future hadron colliders whose experimental
accuracy should be at the few permille level. Actually, the logarithmic contributions to the amplitudes
may reach the few tens of percent level at the high energy which is reached at the LHC and the
validity of the simple one-loop approximation must be seriously
questioned \cite{Beccaria1}. From this point of view, If the high
energy behaviour of the amplitudes for neutralino pair production at
proton-proton collision is considered, one-loop EW
corrections should be kept in view. Since the nonlogarithmic
one-loop contributions come into view to reach at the few percent
level, which is also the level of the expected experimental
accuracy, it may be adequate to disregard these difficult to figure out
effects in the neutralino pair production processes at the LHC energies.

We now present the formulas of the leading and
subleading EW logarithmic corrections for amplitudes of the
subprocess $q\bar{q}\to
\widetilde\chi_{i}^{0}\widetilde\chi_{j}^{0}$, are included in Refs.
\cite{Beccaria1,Beccaria2}. At the one-loop level, these corrections
can be separated into three types of terms as follows:
Renormalization Group (RG) terms, Universal terms and Non-Universal
terms (angular and process dependent terms).

\begin{description}
\item[$(a)$] \textit{Renormalization Group (RG) terms:} The RG contributions represent the linear
logarithms \cite{Degrassi}, which are produced by the running of
the gauge coupling constants, which are known and can be calculated in a
straightforward way. These terms are obtained by introducing in Born
amplitude the running couplings ($g,g^{\prime}$) of the
$SU(2)\otimes U(1)$ according to the asymptotic MSSM
$\beta$-functions are defined as:
\begin{equation} \label{eq:betafunctions}
\widetilde\beta_0=\frac{3}{4}
C_A-\frac{n_g}{2}-\frac{n_h}{8}=-\frac{1}{4} ~~\text{and}
\hspace{0.5cm}
\widetilde\beta_0^{\prime}=-\frac{5}{6}n_g-\frac{n_h}{8}=-\frac{11}{4}
\end{equation}
with
\begin{equation} \label{eq:g^2(s)}
g^2(s)=\frac{g^2(\mu^2)}{1+\widetilde\beta_0
\frac{g^2(\mu^2)}{4\pi^2}\ln(\frac{s}{\mu^2})},\hspace{0.5cm}
g^{\prime 2}(s)=\frac{g^{\prime
2}(\mu^2)}{1+\widetilde\beta_0^\prime \frac{g^{\prime
2}(\mu^2)}{4\pi^2}\ln(\frac{s}{\mu^2})},
\end{equation}

where $C_A=2, n_g=3, n_h=2$ in the MSSM and $g=e/s_W$,
$g^{\prime}=e/c_W$. These terms correspond to the subleading
logarithmic (SL) RG corrections just like in the case of the SM, but
now with the MSSM particle spectrum contributing. At the one-loop, these contributions only appear from higgsino
components $(N_{3i}, N_{4i})$ produced through $Z^0$ exchange in the
$s$-channel for subprocess $q\bar{q}\to
\widetilde\chi_{i}^{0}\widetilde\chi_{j}^{0}$. In that case, they
are written as
\begin{equation} \label{eq:T^RG}
T^{RG}=-\frac{1}{4\pi^2}\left(g^4\widetilde\beta_0\frac{dT_{\hat
s}}{dg^2}+g^{\prime4}\widetilde\beta_0^\prime\frac{dT_{\hat
s}}{dg^{^\prime2}}\right)\ln({\hat s}/\mu^2),
\end{equation}
where $T_{\hat s}$ is the \textit{s}-channel amplitude and $\mu$ is a
reference scale defining the numerical values of $g$, $g^\prime$.
Applying this procedure to the amplitudes, by means of the
substitutions are given as;
\begin{equation} \label{eq:RGT1}
\frac{e^2L_q}{s^2_Wc^2_W}\rightarrow-\frac{2I^3_{q_L}}{4\pi^2}
\left(g^4\widetilde\beta_0+g^{\prime4}\widetilde\beta_0^\prime\left[1-2|Q_q|\right]\right)\ln({\hat
s}/\mu^2),
\end{equation}
\begin{equation} \label{eq:RGT2}
\frac{e^2R_q}{s^2_Wc^2_W}\rightarrow\frac{2Q_q}{4\pi^2}\left(g^{\prime4}\widetilde\beta_0^\prime\right)\ln({\hat
s}/\mu^2).
\end{equation}

\item[$(b)$] \textit{Universal electroweak (EW) terms:} These are process-independent terms, which appear as correction factors to the Born amplitude. Also called ``Sudakov'' terms, these terms
appear to be typically of the form $\left[2\ln(\frac{\hat
s}{m_W^2})-\ln^2(\frac{\hat s}{m_W^2})\right]$ and in a covariant
gauge are generated by diagrams of vertex (initial/final
triangles) and of box type. They are specific of the quantum numbers
and chirality of each external particle line and consist of
``\textsl{Yukawa}'' and ``\textsl{gauge}'' contributions associated
to this line. In addition, they depend on the type of interaction
and on the energy. The universal EW terms appearing in
$q\bar{q}\to \widetilde\chi_{i}^{0}\widetilde\chi_{j}^{0}$ can be
separated into two group: the contributions associated with external
quark (initial) and neutralino (final) lines as given:
\\
\hspace*{1.0cm}\underline{External quark line of chirality $a=L,R$:}
The quark lines correspond to a definite chirality $a$, since all quarks other than third family quarks are taken as
massless as far as the kinematics are concerned. The sum of amplitudes for
the subprocess $q\bar{q}\to
\widetilde\chi_{i}^{0}\widetilde\chi_{j}^{0}$ is $T_{a}^{ij}$ which
is defined by adding indices ($a,i,j$) to
Eq.~\eqref{eq:matrixelement}. The contribution from external quark
line of chirality to $T_{a}^{ij}$ is written as
\begin{equation} \label{eq:Tc(qq)}
T_{a}^{ij}\cdot\left(c_{a}^{q\bar{q}}\right),
\end{equation}
where ${a}$-index refers to exchanged $q_L$ and $q_R$ quarks, and
($i,j$) describes type of the final neutralinos. The factor in
Eq.~\eqref{eq:Tc(qq)}, $c_{a}^{q\bar{q}}$ is given as
\begin{equation} \label{eq:c(q)a}
c_{a}^{q\bar{q}}=c_{gauge,a}^{q\bar{q}}+c_{Yukawa,a}^{q\bar{q}},
\end{equation}
where the gauge term is
\begin{equation} \label{eq:c(qgauge)}
c_{gauge,a}^{q\bar{q}}=\frac{\alpha}{8\pi}\left[\frac{I_{q_a}(I_{q_a}+1)}{s_W^2}+\frac{Y_{q_a}^2}{4c_W^2}\right]\left[2\ln(\frac{\hat
s}{m_W^2})-\ln^2(\frac{\hat s}{m_W^2})\right],
\end{equation}
while the Yukawa term is defined as
\begin{equation} \label{eq:c(qyuk)}
\begin{split}
c_{Yukawa,a}^{q\bar{q}} =&
-\frac{\alpha}{16\pi{s_W^2}}\left[\ln(\frac{\hat
s}{m_W^2})\right]\biggl\{\left[\frac{m_t^2}{m_W^2s_\beta^2}+\frac{m_b^2}{m_W^2c_\beta^2}\right]\delta_{aL} \\
  & +2\left[\frac{m_t^2}{m_W^2s_\beta^2}\delta_{I_{q_a}^{3},1/2}+
  \frac{m_b^2}{m_W^2c_\beta^2}\delta_{I_{q_a}^{3},-1/2}\right]\delta_{aR}\biggr\}
\end{split}
\end{equation}

and only this term appears for bottom and top quarks since masses of
the other quarks can be neglected. In Eq.~\eqref{eq:c(qgauge)},
$I_{q_a}$ is the full weak isospin of the quark with chirality $a$,
and $Y_{q_a}$ is the hypercharge which is defined as
$Y_{q_a}=2(Q_{q_a}-I_{q_a}^3)$. Consequently, the amplitude of the
$q\bar{q}\to \widetilde\chi_{i}^{0}\widetilde\chi_{j}^{0}$ with this
contribution can be written as
\begin{equation} \label{eq:T(1loop)}
T_{\textrm{one-loop}}^{ij}=\left[1+c_{a}^{q\bar{q}}\right]
T_{a}^{ij}.
\end{equation}

\hspace*{1.0cm}\underline{External neutralino line of chirality
$b=L,R$ :} The contribution from external neutralino line of
chirality to $T_{b}^{ij}$ may be written as
\begin{equation} \label{eq:Tc(NiNj)}
\sum_{k}\left[T_{b}^{ik}\cdot{c_{b}^{\widetilde{\chi}_{k}^{0}\widetilde{\chi}_{j}^{0}}}+T_{b}^{kj}\cdot{c_{b}^{\widetilde{\chi}_{k}^{0}\widetilde{\chi}_{i}^{0}*}}\right].
\end{equation}
Here, one use a matrix notation for external particle is one member
of mixed states. The amplitude of the subprocess $q\bar{q}\to
\widetilde\chi_{i}^{0}\widetilde\chi_{j}^{0}$ involves the neutral
higgsino components $(N_{3i}, N_{4i})$ produced through $Z^0$
exchange in the $s$-channel, but it only involves the neutral
gaugino $(\widetilde W_3)$ component $(N_{2i})$ produced through
squark exchange in the $t$- and $u$-channels. In addition to this,
the logarithmic contributions for higgsino $s$-channel amplitude
($T_{\hat{s}}$ ) involve both the ``\textsl{higgsino, gauge}'' and
``\textsl{higgsino, Yukawa}'' parts, whereas for the gaugino $t$-
and $u$-channels amplitudes ($T_{\hat{t}}$ and $T_{\hat{u}}$), only
include the ``\textsl{gaugino, gauge}'' part. Thus, these
contributions may be written as
\begin{equation} \label{eq:c(NiNj)a}
c_{b}^{\widetilde{\chi}_{i}^{0}\widetilde{\chi}_{j}^{0}}=c_{higgsino,gauge,b}^{\widetilde{\chi}_{i}^{0}\widetilde{\chi}_{j}^{0}}+c_{higgsino,yuk,b}^{\widetilde{\chi}_{i}^{0}\widetilde{\chi}_{j}^{0}}
+c_{gaugino,gauge,b}^{\widetilde{\chi}_{i}^{0}\widetilde{\chi}_{j}^{0}},
\end{equation}
where
\begin{equation} \label{eq:c(higgauge)a}
\begin{split}
c_{higgsino,gauge,b}^{\widetilde{\chi}_{i}^{0}\widetilde{\chi}_{j}^{0}}
=&\frac{\alpha(1+2c_W^2)}{32\pi{s_W^2}{c_W^2}}\left[2\ln(\frac{\hat
s}{m_W^2})-\ln^2(\frac{\hat s}{m_W^2})\right] \\
  &\times\left[(N_{4i}^*N_{4j}+N_{3i}^*N_{3j})\delta_{bL}+(N_{4i}N_{4j}^*+N_{3i}N_{3j}^*)\delta_{bR}\right],
\end{split}
\end{equation}

\begin{equation} \label{eq:c(higyuk)a}
\begin{split}
c_{higgsino,yuk,b}^{\widetilde{\chi}_{i}^{0}\widetilde{\chi}_{j}^{0}}
=&-\frac{3\alpha}{16\pi{s_W^2}{m_W^2}}\left[\ln(\frac{\hat
s}{m_W^2})\right]\biggl\{\frac{m_t^2}{s_\beta^2}(N_{4i}^*N_{4j}\delta_{bL}+N_{4i}N_{4j}^*\delta_{bR}) \\
  & +\frac{m_b^2}{c_\beta^2}(N_{3i}^*N_{3j}\delta_{bL}+N_{3i}N_{3j}^*\delta_{bR})\biggr\}
\end{split}
\end{equation}

and
\begin{equation} \label{eq:c(gauginogauge)a}
c_{gaugino,gauge,b}^{\widetilde{\chi}_{i}^{0}\widetilde{\chi}_{j}^{0}}=-\frac{\alpha}{4\pi{s_W^2}}\left[\ln^2(\frac{\hat
s}{m_W^2})\right]\left[N_{2i}^*N_{2j}P_L+N_{2i}N_{2j}^*P_R\right].
\end{equation}
One sees from these contributions, the $\left[2\ln(\frac{\hat
s}{m_W^2})-\ln^2(\frac{\hat s}{m_W^2})\right]$ combination can also
be found in the higgsino components, and $\left[-\ln^2(\frac{\hat
s}{m_W^2})\right]$ term in the gaugino components. The leads to an
additional potential check of the assumed supersymmetric nature of
the interactions of neutralinos which can be achieved by a
measurement of the production rate of the four neutralinos
\cite{Beccaria3}. Consequently, this contribution to the amplitude
of the $q\bar{q}\to \widetilde\chi_{i}^{0}\widetilde\chi_{j}^{0}$
can be carried out as
\begin{equation} \label{eq:T(1loop)}
T_{\textrm{one-loop}}^{ij}=\sum_{k}\left[\delta_{ij}+\delta_{jk}\cdot{c_{b}^{\widetilde{\chi}_{k}^{0}\widetilde{\chi}_{j}^{0}}}+\delta_{ki}\cdot{c_{b}^{\widetilde{\chi}_{k}^{0}\widetilde{\chi}_{i}^{0}*}}\right]T_{b}^{ij}.
\end{equation}

\item[$(c)$] \textit{Angular and process dependent terms:} They only consist in residual terms arising from the quadratic logarithms $ln^2t$ or $ln^2u$
produced by box diagrams containing $ Z^0, W^\pm$ and $\gamma$ gauge
boson internal lines, where $t=-\frac{s}{2}(1-cos(\theta))$ and
$u=-\frac{s}{2}(1+cos(\theta))$, $\theta$ being the scattering
angle. There are only few such diagrams and they have been all
clearly calculated. The diagrams with internal $Z$ lines can be disregarded,
because their contributions become orthogonal to the
Born terms and cannot interfere with them.
\end{description}
For amplitude of the each subprocess, we can write,
\begin{equation} \label{eq:T(1loop ew)}
T_{\textrm{One-loop EW}}^{ij}=\left[1+c_{a}^{q\bar{q}\rightarrow
\widetilde{\chi}_{i}^{0}\widetilde{\chi}_{j}^{0}}\right] T_{a}^{ij},
\end{equation}
where $c_{a}^{q\bar{q}\rightarrow
\widetilde{\chi}_{i}^{0}\widetilde{\chi}_{j}^{0}}$ includes all of
the contributions given above. We have exactly calculated these
three types of contributions for SL logarithmic accuracy. The total cross-section including
the EW corrections reads
\begin{equation} \label{eq:s(1loop ew)}
\sigma=\sigma_{0}+\Delta\sigma=\sigma_{0}(1+\delta),
\end{equation}
where $\sigma_{0}$ is the Born level cross-section, $\Delta\sigma$ is the full electroweak
contribution to cross-section and $\delta$ is the EW relative correction.

\newpage
\section{Numerical results and discussion}\label{results}

In this section, we present a detailed numerical study of the neutralino pair production process $pp \to q \bar{q}
\to \widetilde{\chi}_{i}^{0}\widetilde{\chi}_{j}^{0}$ at the LHC
energies with special emphasis on effects of the EW logarithmic contributions, which are so
important thereby can reach the few tens of percent level at the
high energy. Focusing on the lightest neutralino
$\widetilde{\chi}_{1}^{0}$ is likely to be the LSP and the
next-to-lightest neutralino $\widetilde{\chi}_{2}^{0}$, we investigate the
relevant processes $pp \to \widetilde
{\chi}_{1}^{0}\widetilde{\chi}_{1}^{0}$, $pp \to \widetilde
{\chi}_{2}^{0}\widetilde{\chi}_{2}^{0}$ and $pp \to \widetilde
{\chi}_{1}^{0}\widetilde{\chi}_{2}^{0}$, can be the most dominant
neutralino pair production processes. In our numerical calculations, we just limit the values
of $M_1$, $M_2$ and $\mu$ to be real and positive, and
we set $\tan\beta=45$,
$m_{\widetilde{u}_{L}}$= 998.56 GeV, $m_{\widetilde{u}_{R}}$= 999.36 GeV, $m_{\widetilde{d}_{L}} $= 1000.31 GeV, $m_{\widetilde{d}_{R}}$= 1001.77 GeV. In addition,
we fix the chargino masses as $m_{\widetilde{\chi}_{1}^{+}}=85.99$
GeV and $m_{\widetilde{\chi}_{2}^{+}}=206.00$ GeV for higgsino and gaugino-like scenarios, and $m_{\widetilde{\chi}_{1}^{+}}=87.89$
GeV and $m_{\widetilde{\chi}_{2}^{+}}=204.09$ GeV for mixture-case. When using
Eqs.~\eqref{eq:2M2} and \eqref{eq:2mu2} with given chargino masses,
there appear three different cases to choices of the parameters
$\mu$ and $M_2$, as mentioned previously, these are the
higgsino-like, the gaugino-like and mixture-case respectively.

\begin{figure}[hpt]
    \begin{center}
\includegraphics[scale=0.36]{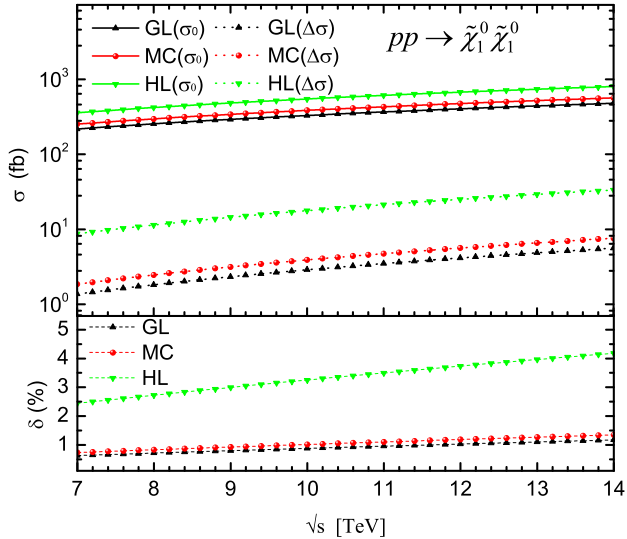}
     \end{center}
\caption{The cross-sections of the process
$pp\to\widetilde{\chi}_{1}^0\widetilde{\chi}_{1}^0$ at
tree level, the EW corrections and the relative corrections
as a function of the center-of-mass energy $\sqrt{s}$.} \label{Fig2}
\end{figure}

\begin{figure}[hpt]
    \begin{center}
\includegraphics[scale=0.36]{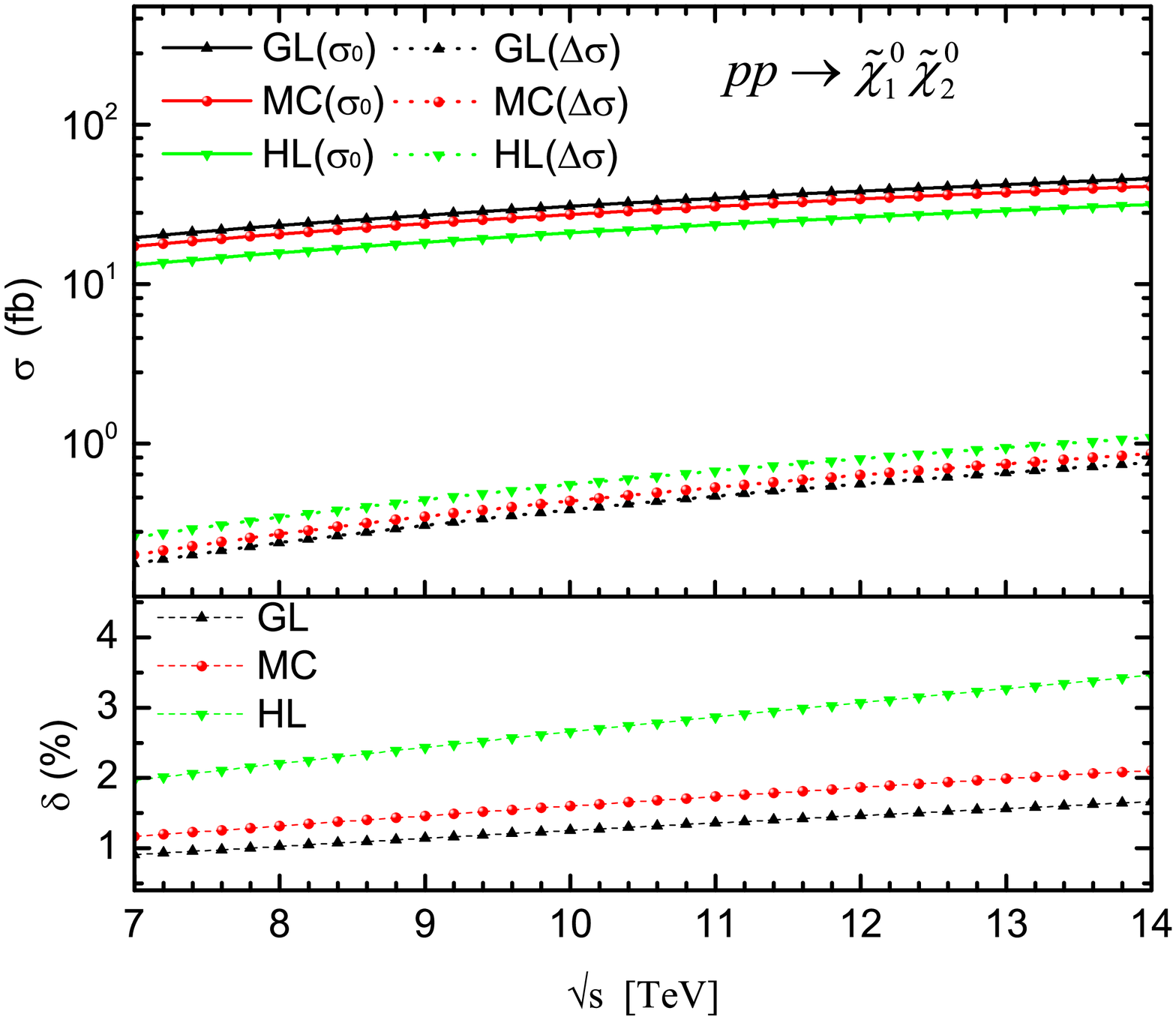}
     \end{center}
\caption{The cross-sections of the process
$pp\to\widetilde{\chi}_{1}^0\widetilde{\chi}_{2}^0$ at
tree level, the EW corrections and the relative corrections
as a function of the center-of-mass energy $\sqrt{s}$.} \label{Fig3}
\end{figure}

\begin{figure}[hpt]
    \begin{center}
\includegraphics[scale=0.36]{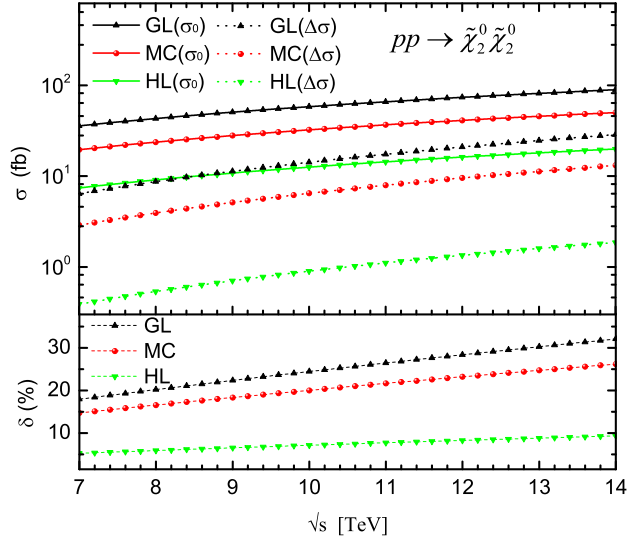}
     \end{center}
\caption{The cross-sections of the process
$pp\to\widetilde{\chi}_{2}^0\widetilde{\chi}_{2}^0$ at
tree level, the EW corrections and the relative corrections
as a function of the center-of-mass energy $\sqrt{s}$.} \label{Fig4}
\end{figure}

\begin{figure}[h]
\begin{center}
\mbox{\epsfig{figure=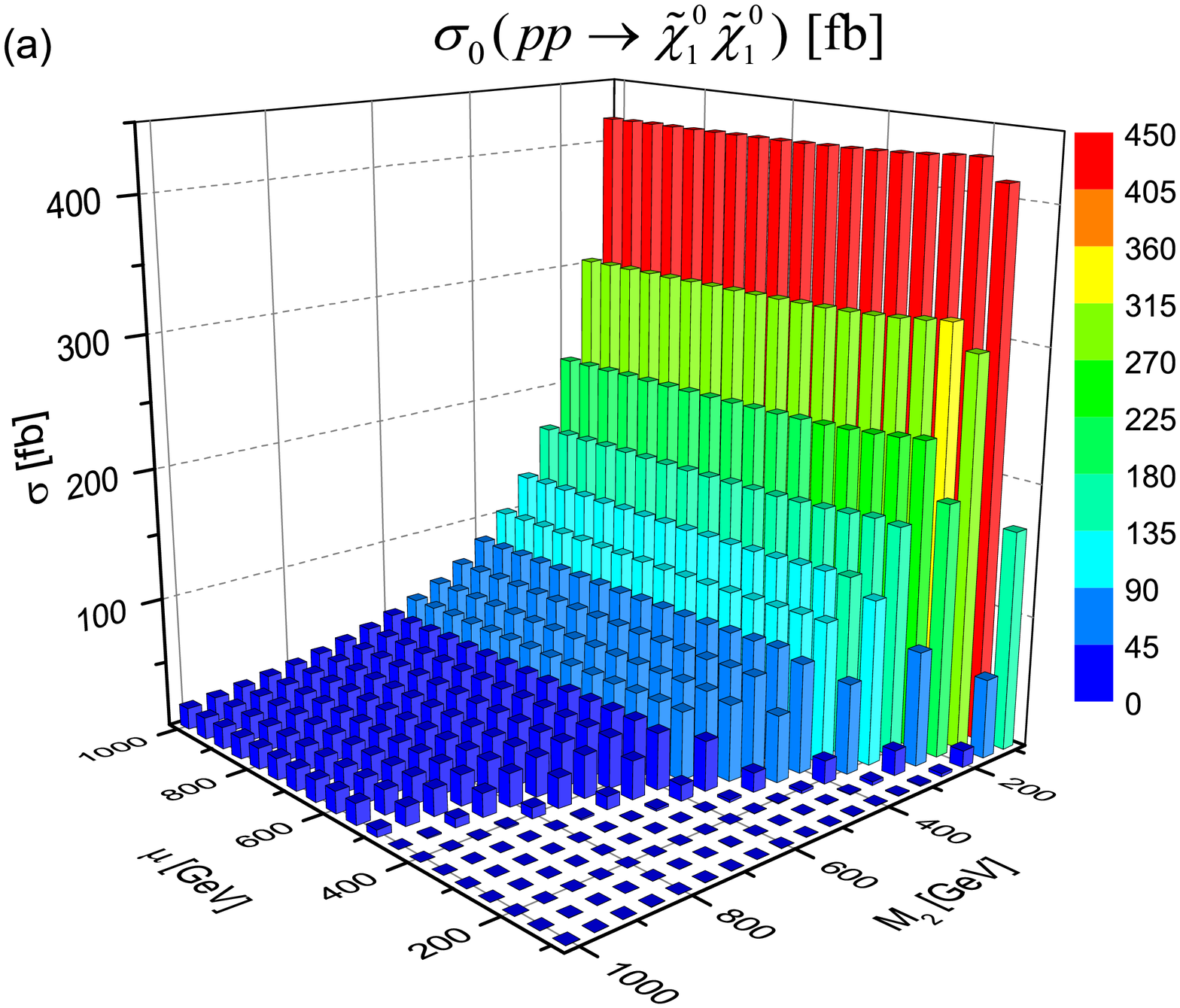,height=7.3cm,width=8.1cm}}
\mbox{\epsfig{figure=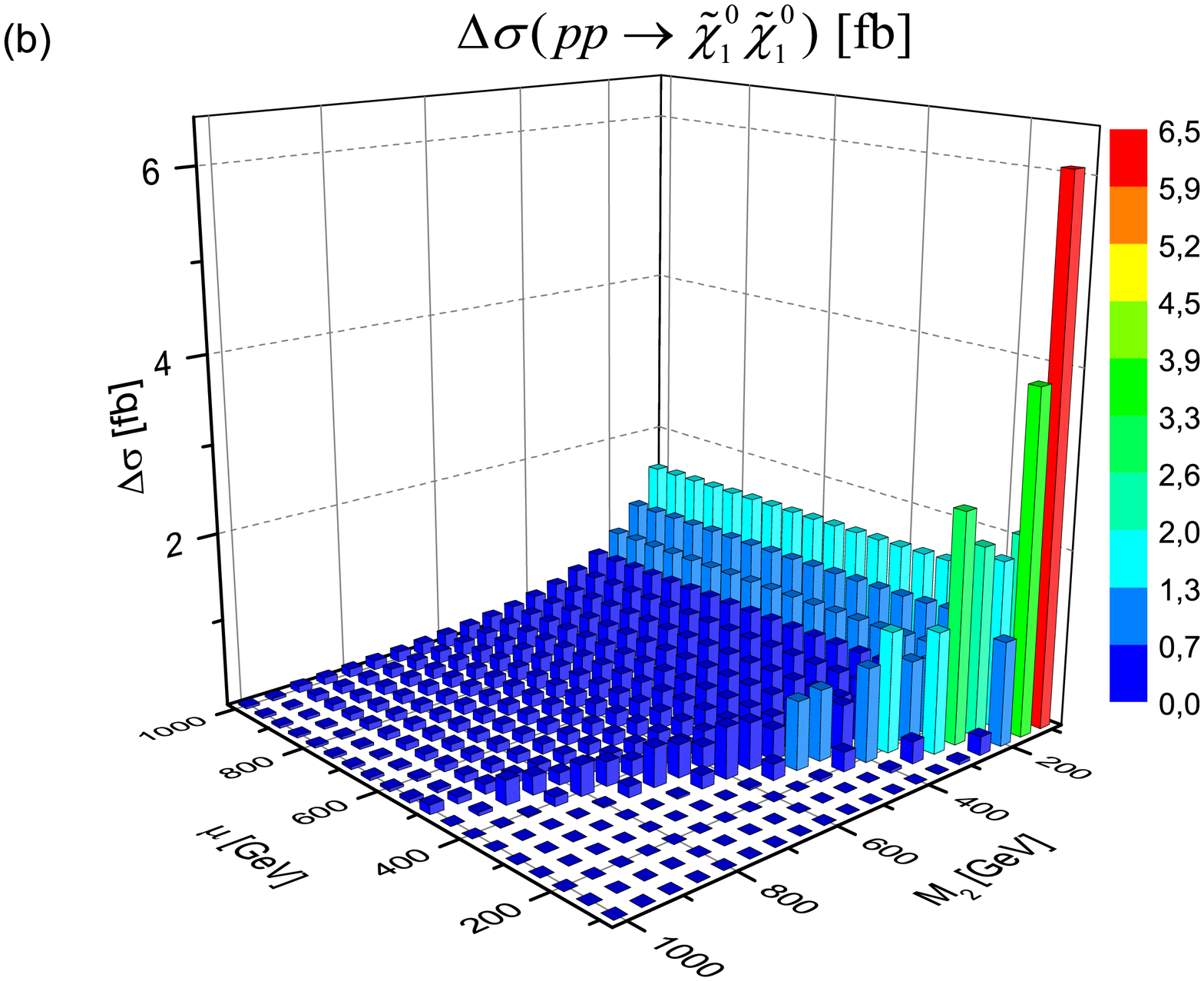,height=7.3cm,width=8.1cm}}
\mbox{\epsfig{figure=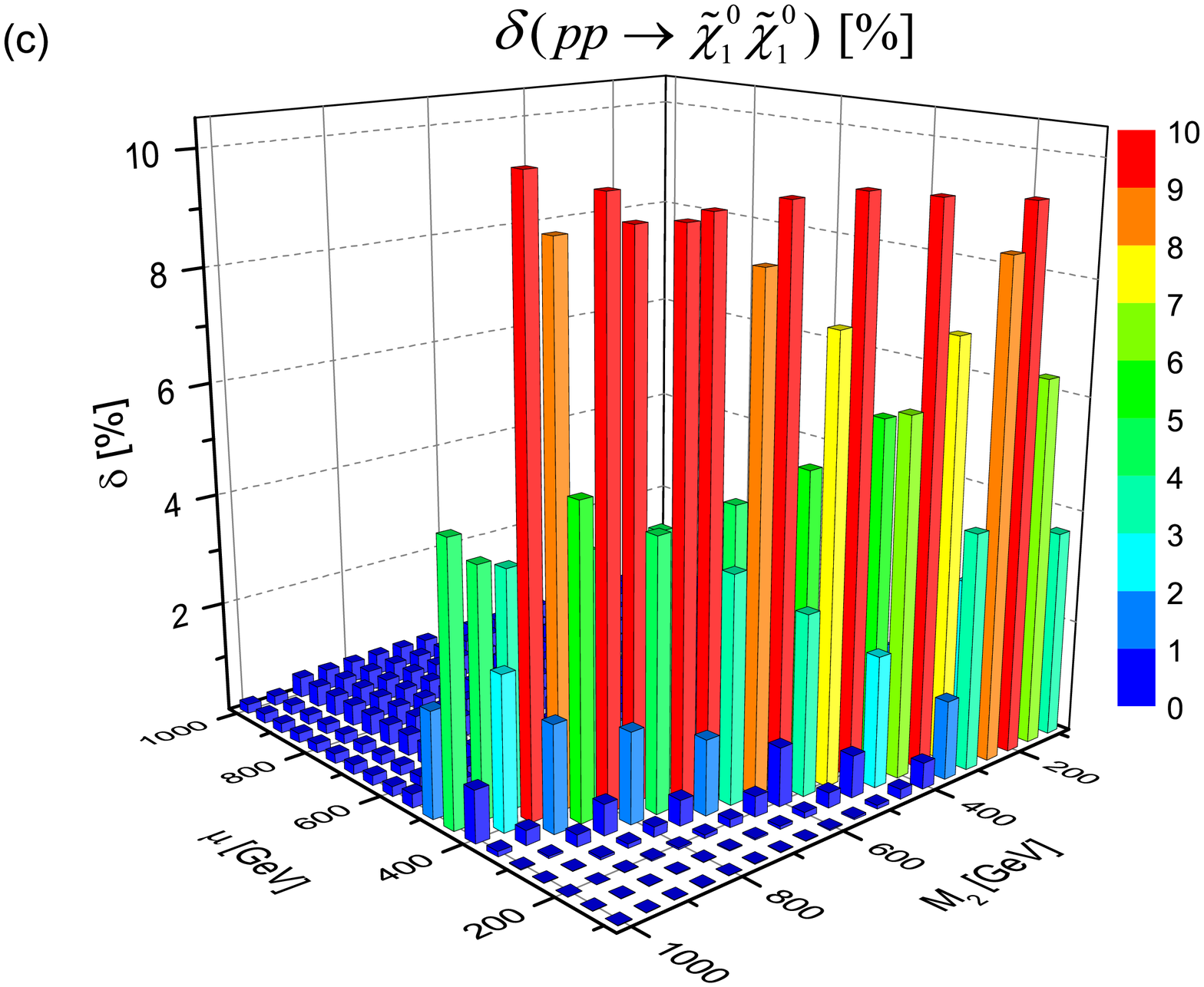,height=7.3cm,width=8.1cm}}
\end{center}
\caption{The cross-section of the process
$p\bar{p}\to\widetilde{\chi}_{1}^{0}\widetilde{\chi}_{1}^0$ (a) at
tree level, (b) the EW correction and (c) the relative correction
as functions of $M_2$ and $\mu$ for $\sqrt{s} =8$ TeV.}
\label{Fig5}
\end{figure}

\begin{figure}[htb]
\begin{center}
\mbox{\epsfig{figure=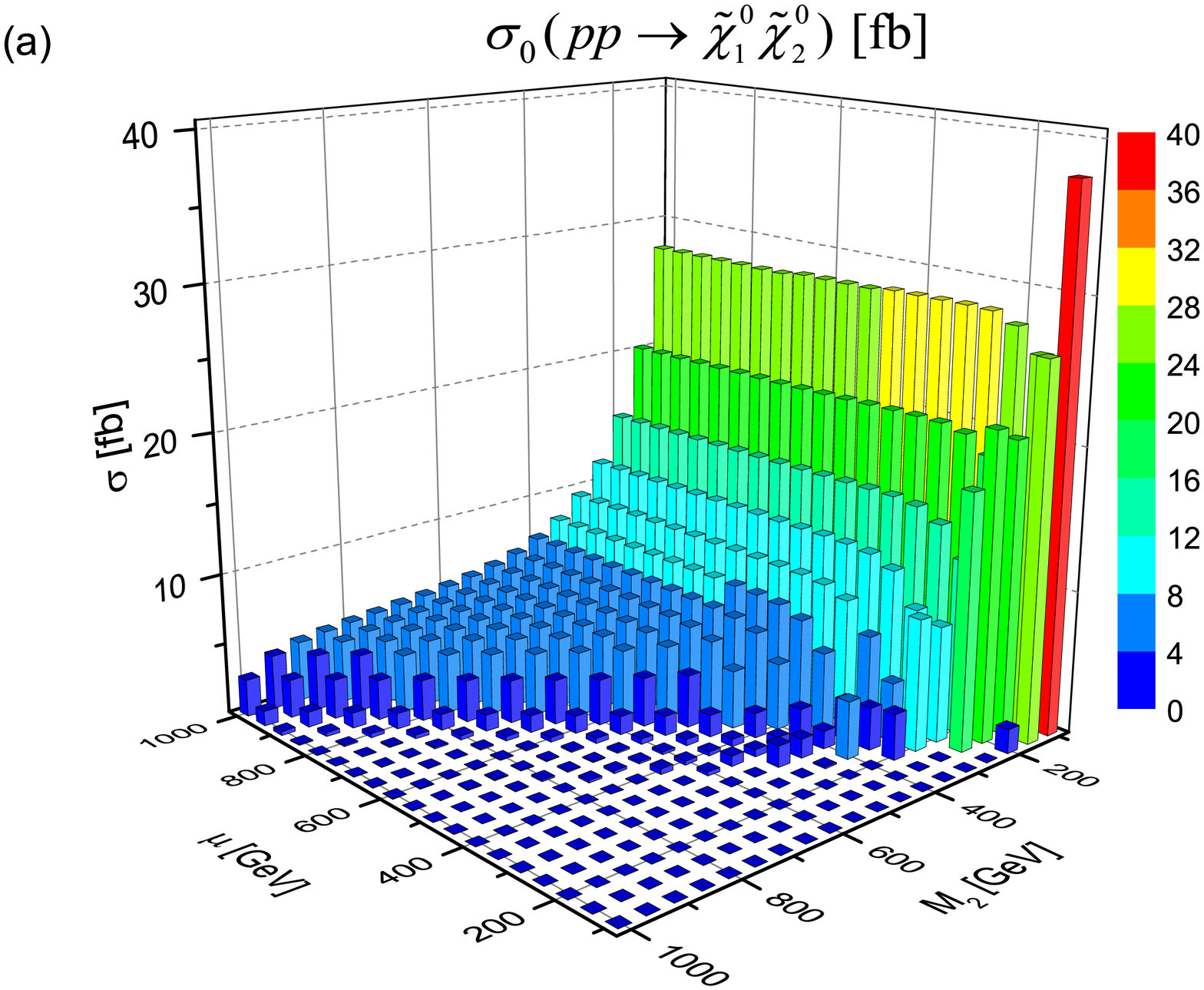
                             ,height=7.3cm,width=8.1cm}}
\mbox{\epsfig{figure=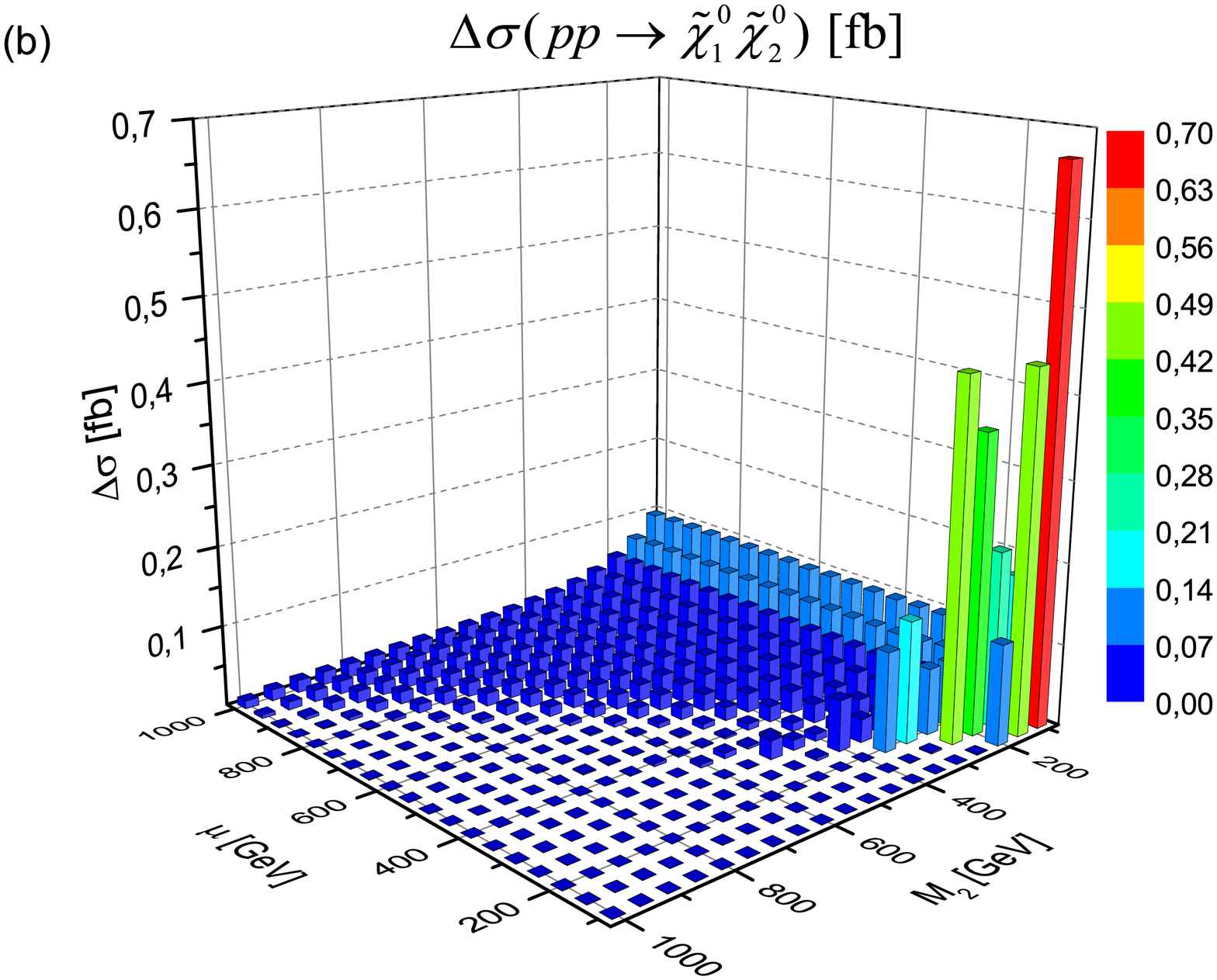
                             ,height=7.3cm,width=8.1cm}}
\mbox{\epsfig{figure=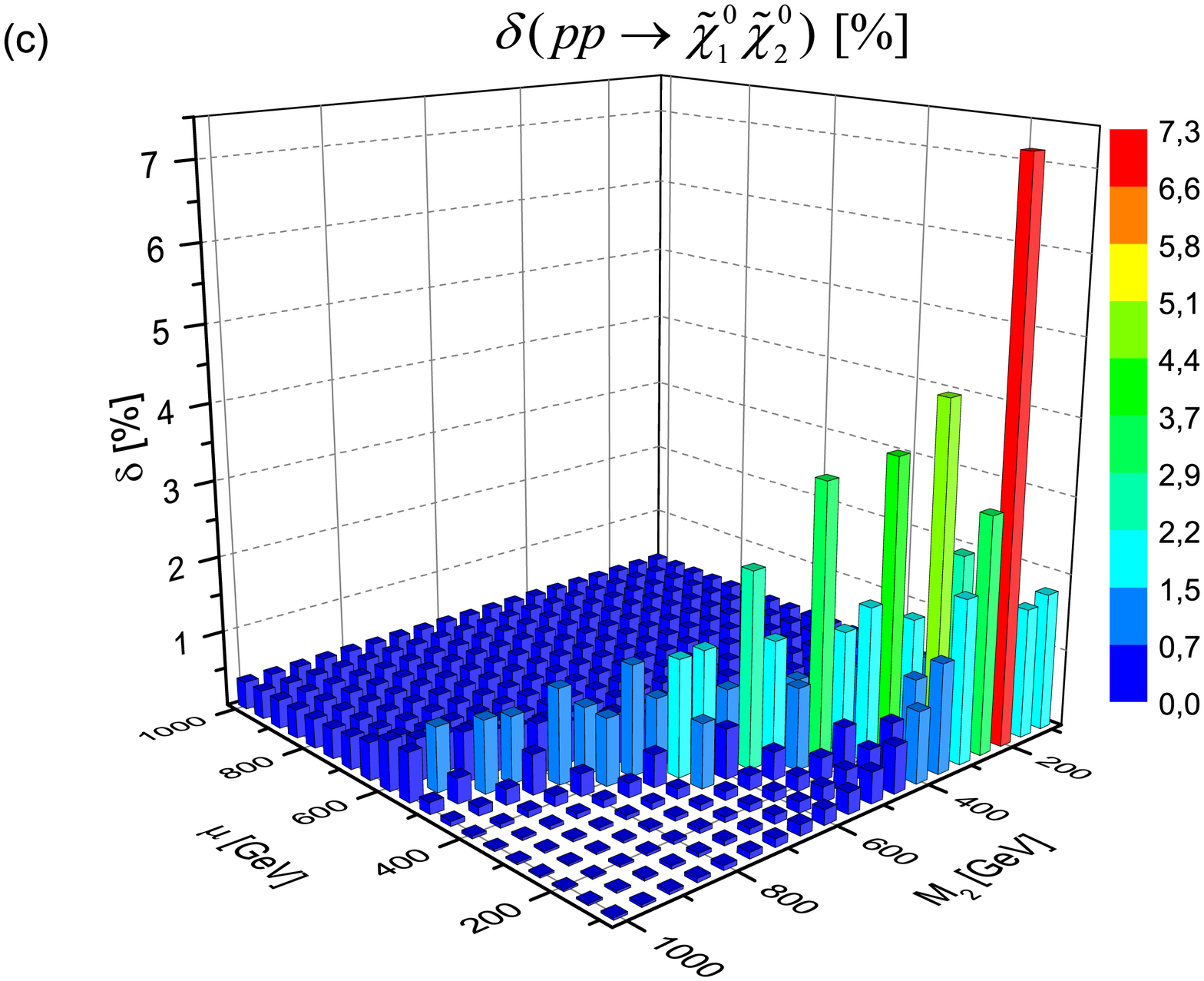
                             ,height=7.3cm,width=8.1cm}}
\end{center}
\caption{The cross-section of the process
$p\bar{p}\to\widetilde{\chi}_{1}^{0}\widetilde{\chi}_{2}^0$ (a) at
tree level, (b) the EW correction and (c) the relative correction
as functions of $M_2$ and $\mu$ for $\sqrt{s} =8$ TeV.}
\label{Fig6}
\end{figure}

\begin{figure}[htb]
\begin{center}
\mbox{\epsfig{figure=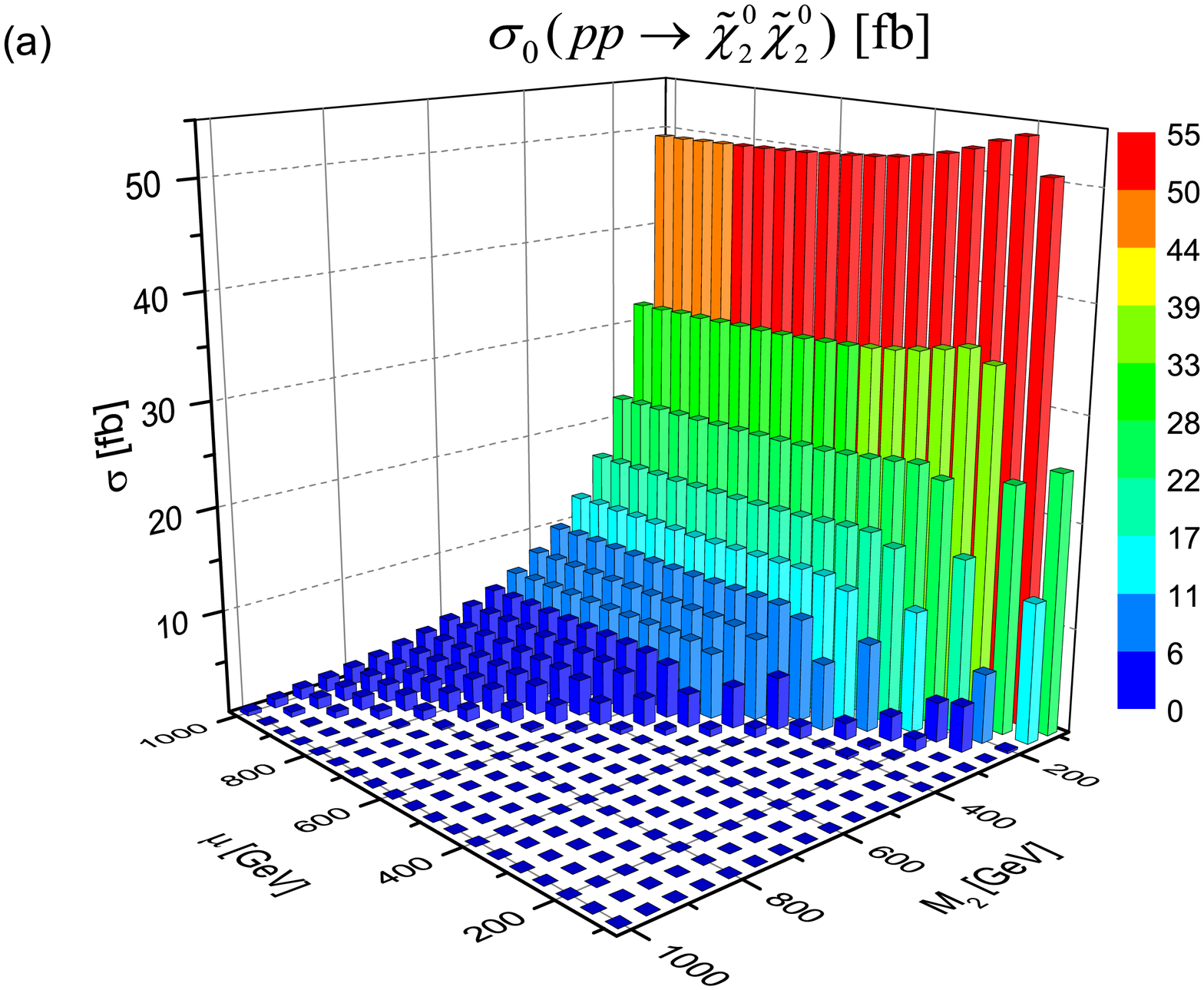
                             ,height=7.3cm,width=8.1cm}}
\mbox{\epsfig{figure=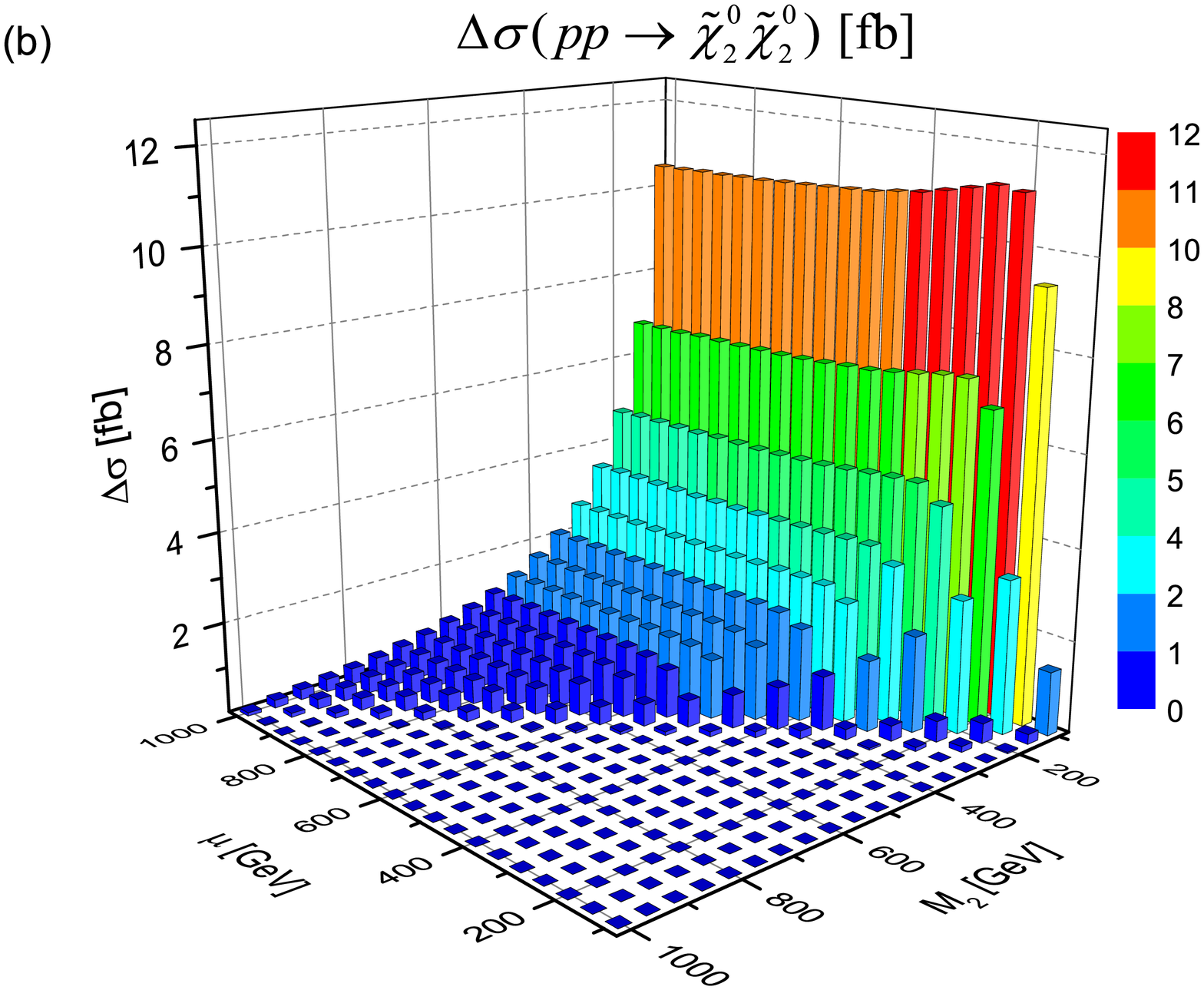
                             ,height=7.3cm,width=8.1cm}}
\mbox{\epsfig{figure=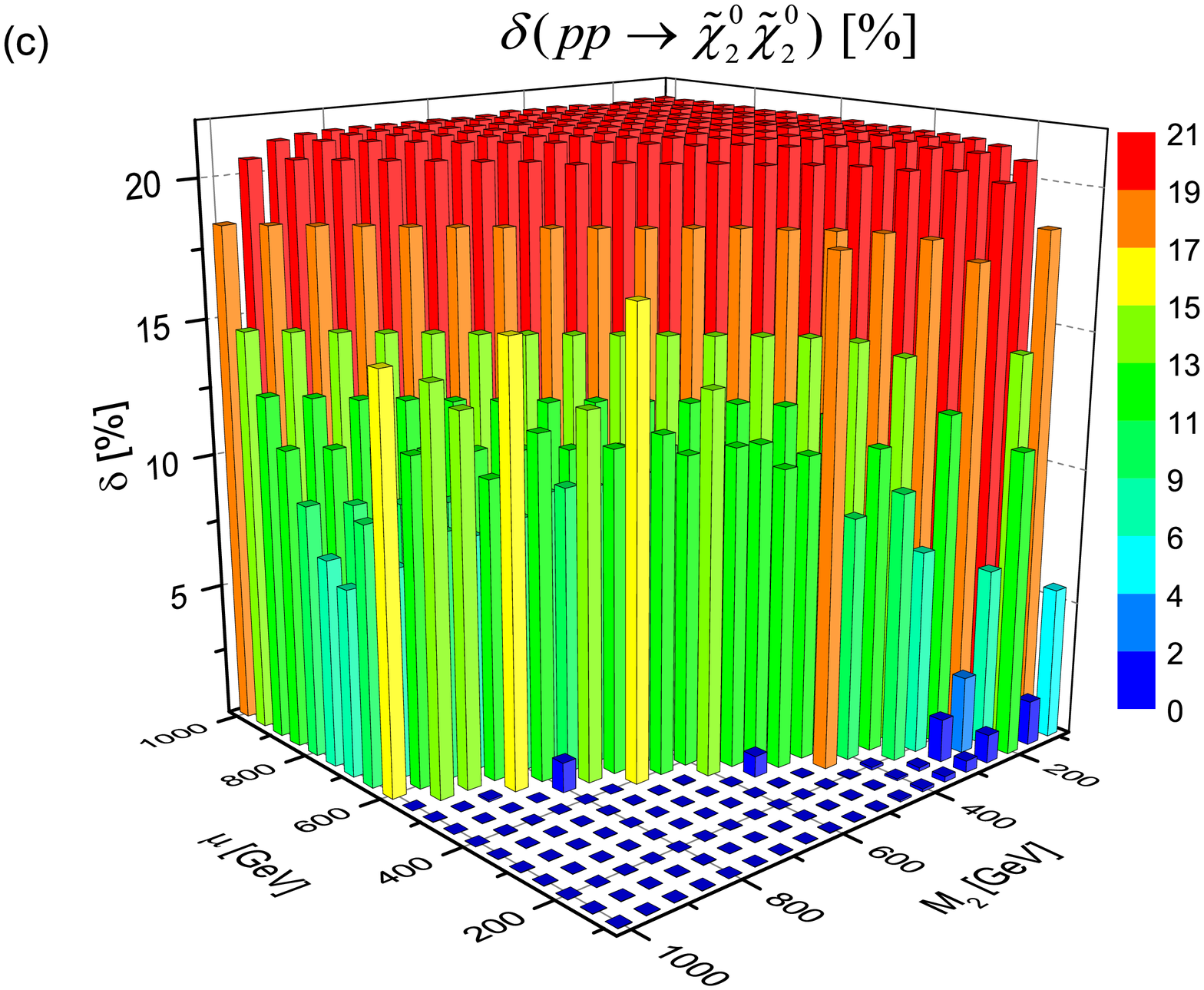
                             ,height=7.3cm,width=8.1cm}}
\end{center}
\caption{The cross-section of the process
$pp\to\widetilde{\chi}_{2}^{0}\widetilde{\chi}_{2}^0$ (a) at
tree level, (b) the EW correction and (c) the relative correction
as functions of $M_2$ and $\mu$ for $\sqrt{s} =8$ TeV.}
\label{Fig7}
\end{figure}

\begin{figure}[hpt]
    \begin{center}
\includegraphics[scale=0.36]{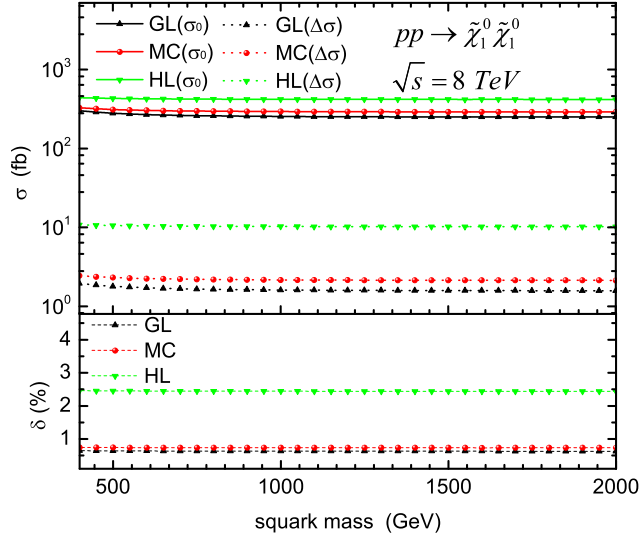}
     \end{center}
\caption{The cross-sections of the process
$pp\to\widetilde{\chi}_{1}^{0}\widetilde{\chi}_{1}^0$ at
tree level, the EW corrections and the relative corrections
as a function of the squark mass at center-of-mass energy
$\sqrt{s} =8$ TeV.} \label{Fig8}
\end{figure}

\begin{figure}[hpt]
    \begin{center}
\includegraphics[scale=0.36]{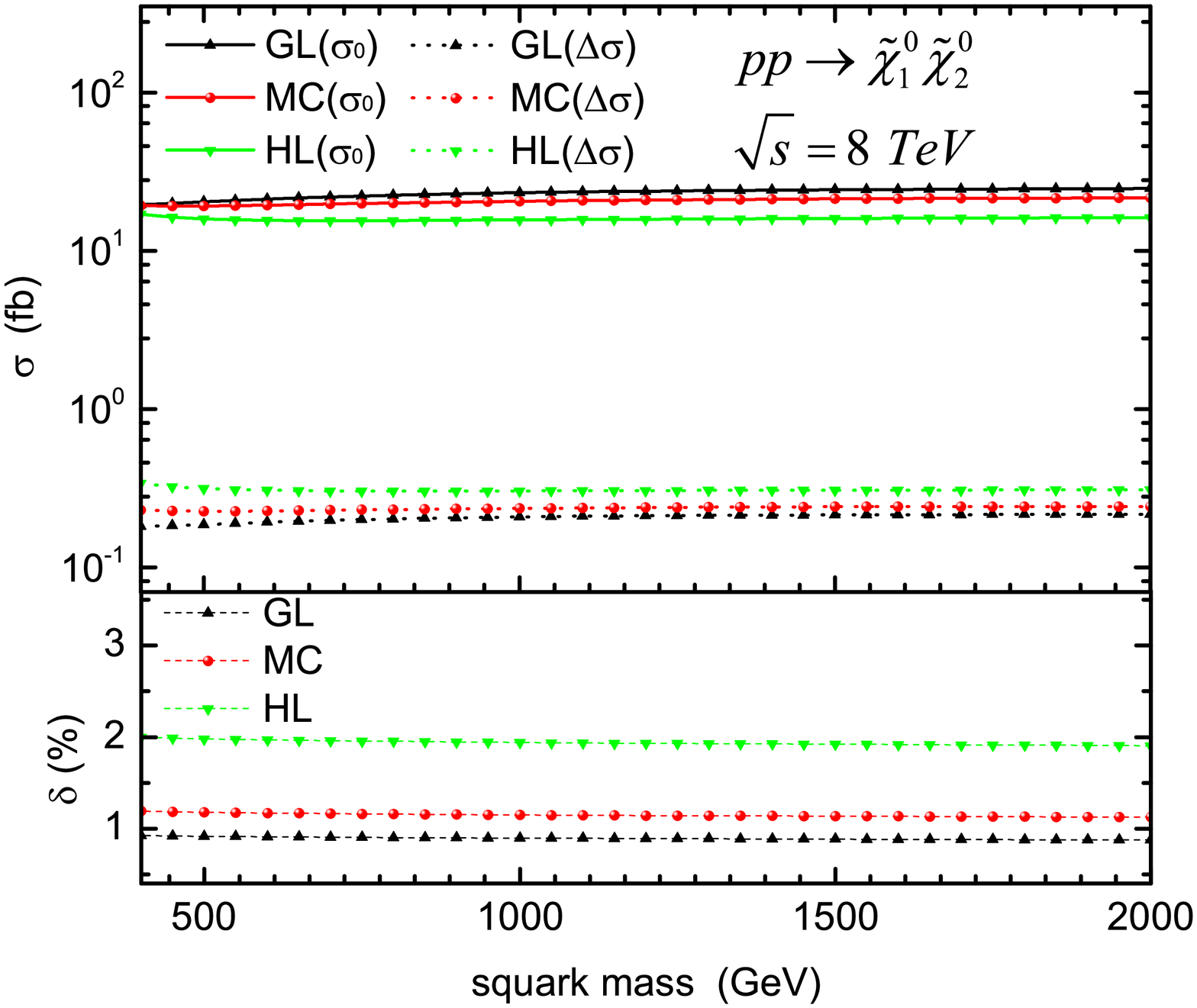}
     \end{center}
\caption{The cross-sections of the process
$pp\to\widetilde{\chi}_{1}^{0}\widetilde{\chi}_{2}^0$ at
tree level, the EW corrections and the relative corrections
as a function of the squark mass at center-of-mass energy
$\sqrt{s} =8$ TeV.} \label{Fig9}
\end{figure}

\begin{figure}[hpt]
    \begin{center}
\includegraphics[scale=0.36]{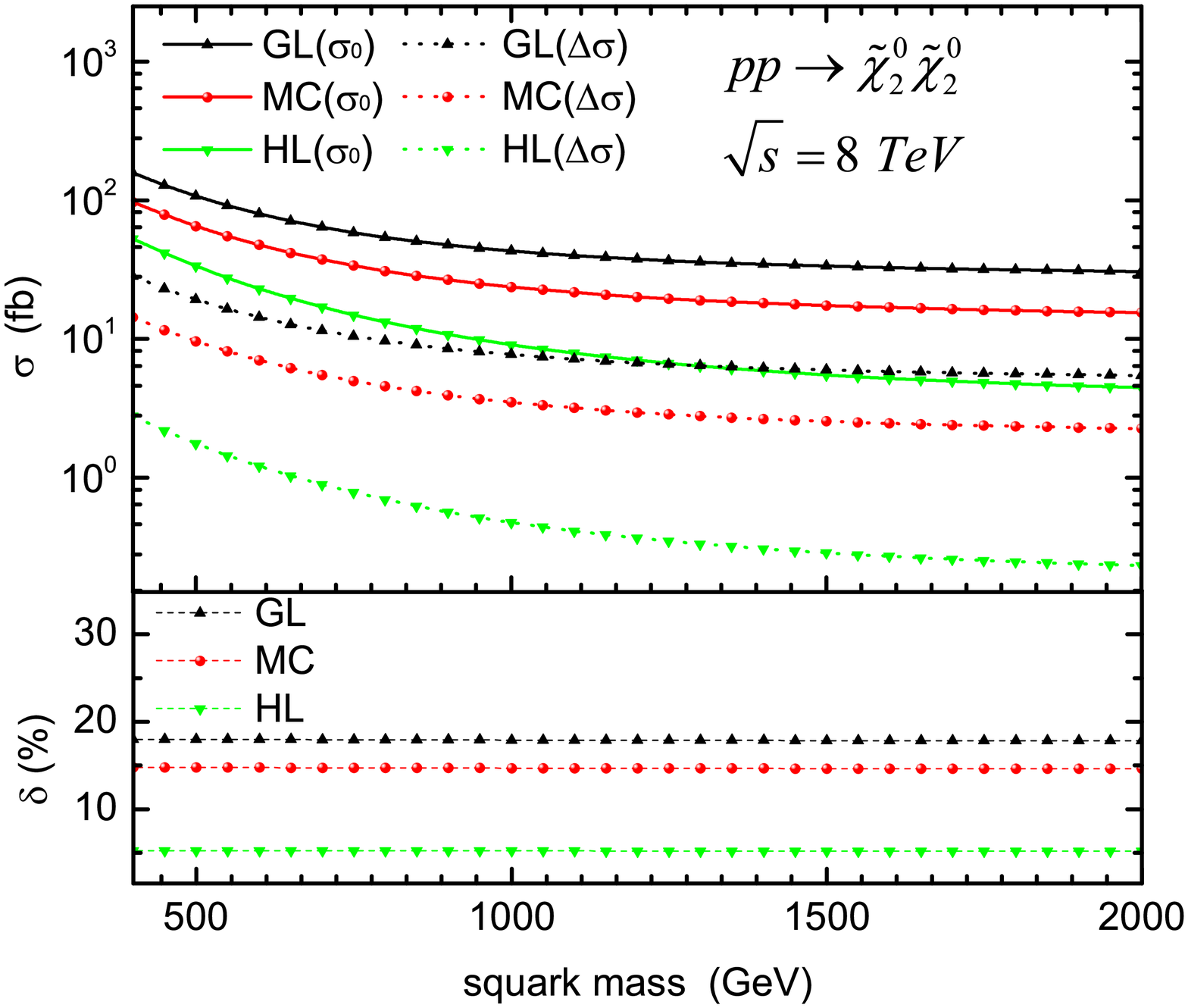}
     \end{center}
\caption{The cross-sections of the process
$pp\to\widetilde{\chi}_{2}^{0}\widetilde{\chi}_{2}^0$ at
tree level, the EW corrections and the relative corrections
as a function of the squark mass at center-of-mass energy
$\sqrt{s} =8$ TeV.} \label{Fig10}
\end{figure}

\begin{figure}[hpt]
    \begin{center}
\includegraphics[scale=0.36]{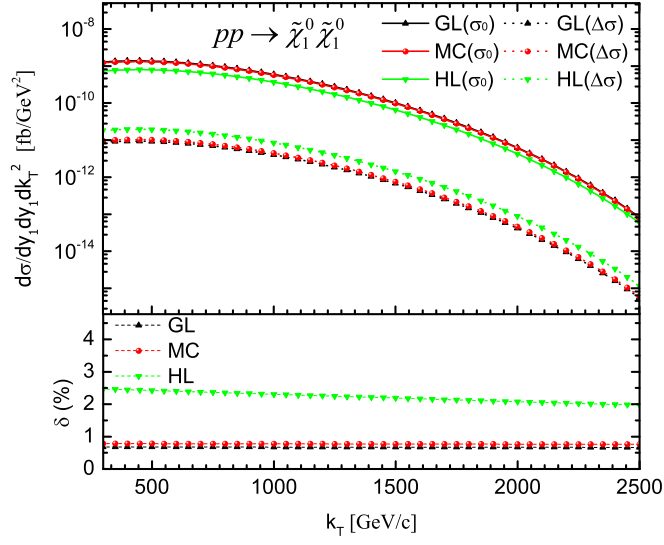}
     \end{center}
\caption{The differential cross-sections of the process $pp \to
\widetilde{\chi}_{1}^0\widetilde{\chi}_{1}^0$ at tree level,
the EW corrections and the relative corrections as a
function of the neutralino pair transverse momentum $k_T$ at
center-of-mass energy $\sqrt{s} =7$ TeV.} \label{Fig11}
\end{figure}

\begin{figure}[hpt]
    \begin{center}
\includegraphics[scale=0.36]{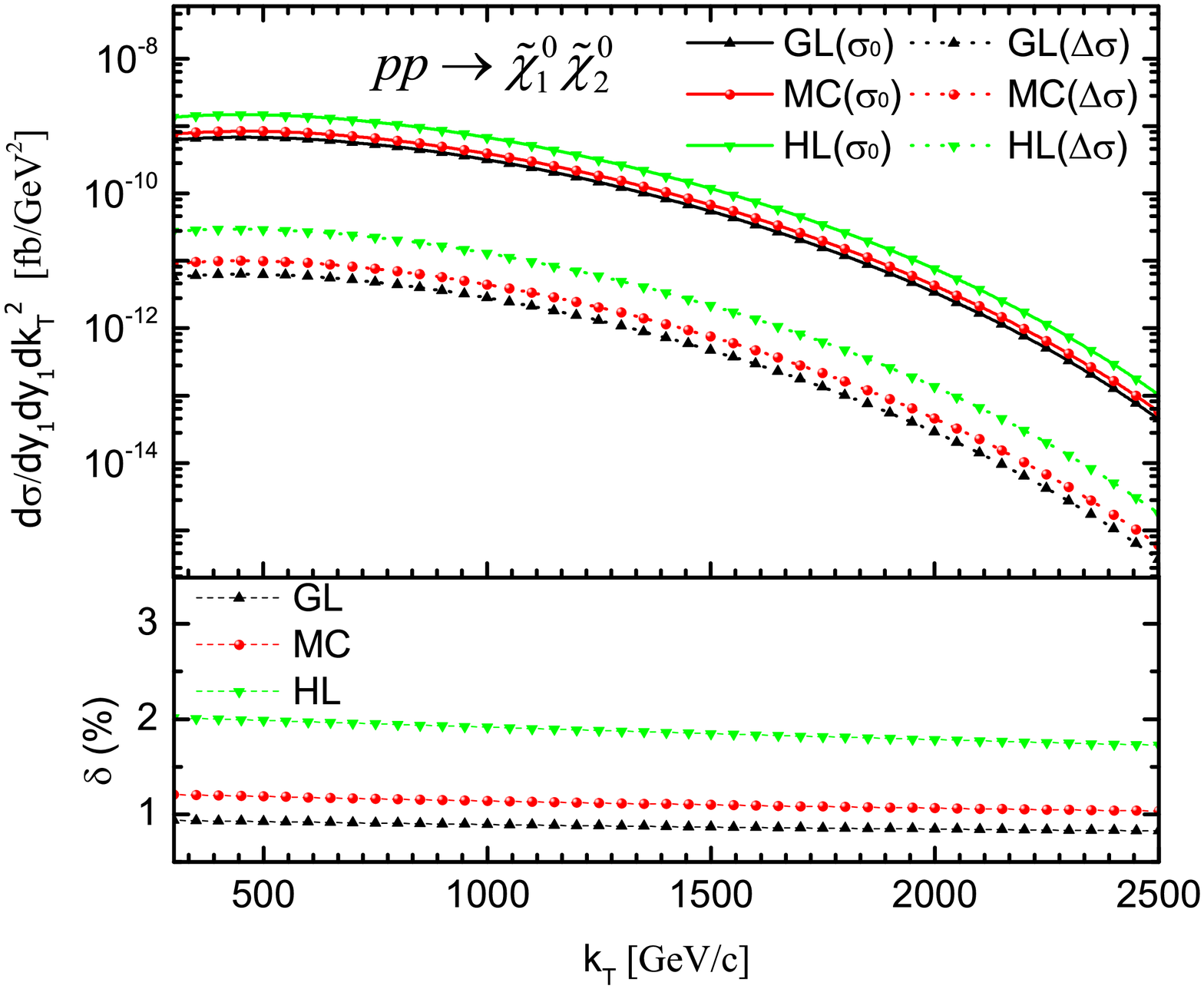}
     \end{center}
\caption{The differential cross-sections of the process $pp \to
\widetilde{\chi}_{1}^0\widetilde{\chi}_{2}^0$ at tree level,
the EW corrections and the relative corrections as a
function of the neutralino pair transverse momentum $k_T$ at
center-of-mass energy $\sqrt{s} =7$ TeV.} \label{Fig12}
\end{figure}

\begin{figure}[hpt]
    \begin{center}
\includegraphics[scale=0.36]{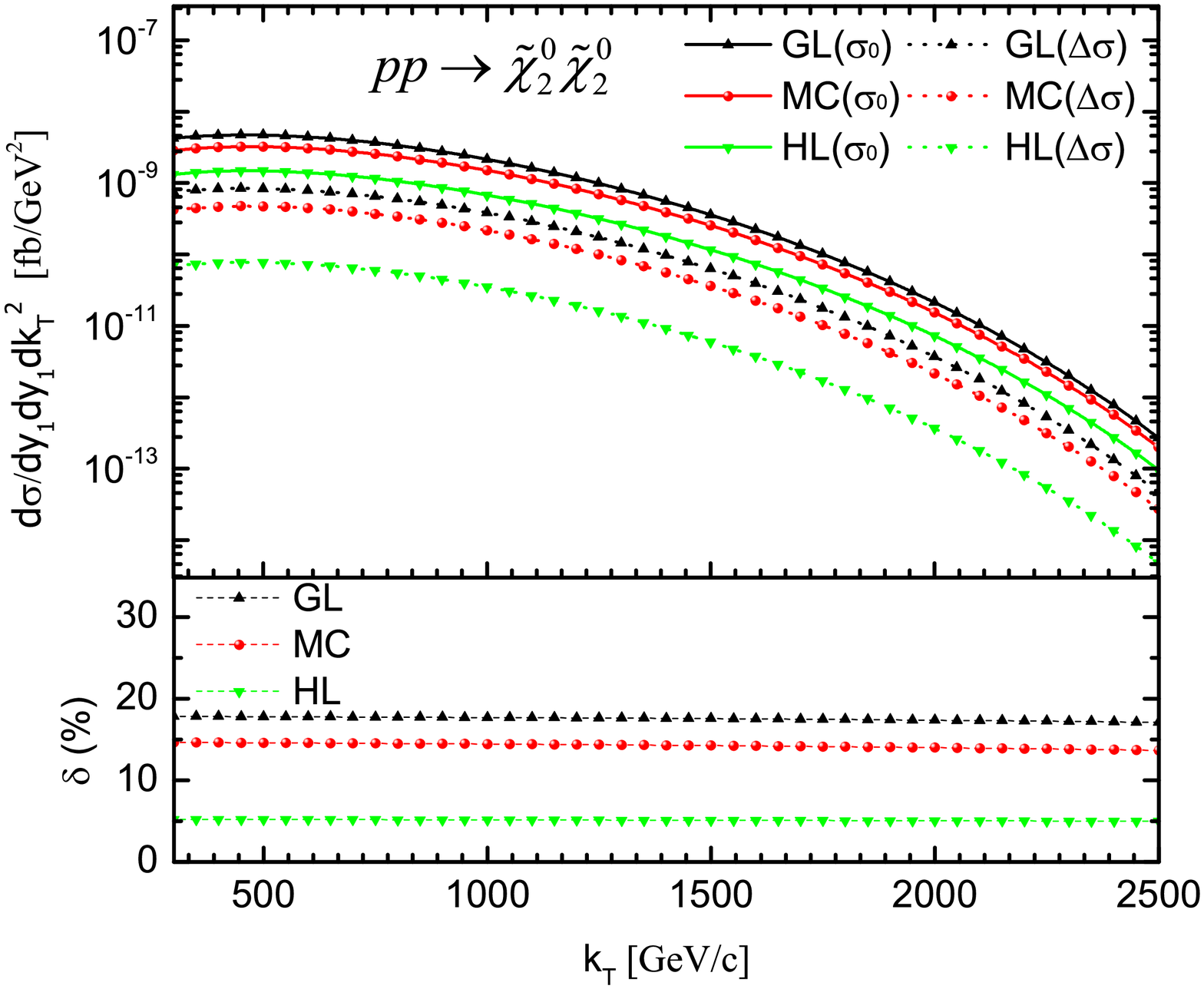}
     \end{center}
\caption{The differential cross-sections of the process $pp \to
\widetilde{\chi}_{2}^0\widetilde{\chi}_{2}^0$ at tree level, the EW corrections
and the relative corrections, as a function of the neutralino pair transverse momentum $k_T$ at
center-of-mass energy $\sqrt{s} =7$ TeV.} \label{Fig13}
\end{figure}

\begin{itemize}
    \item In the higgsino-like case, we obtain $M_2=150$ GeV, $\mu=120$ GeV,
    $M_1=75.309$ GeV and by inserting the values of $M_2$, $\mu$ and
    $M_1$ into Eq.~\eqref{eq:mN}, the neutralino masses are obtained by
$$
m_{\widetilde{\chi}_{1}^{0}}=57.45~\text{GeV}, m_{\widetilde{\chi}_{2}^{0}}=99.00~\text{GeV}, m_{\widetilde{\chi}_{3}^{0}}=136.05~\text{GeV},
m_{\widetilde{\chi}_{4}^{0}}=204.91~\text{GeV}.
$$
    \item In the gaugino-like case, we have $M_2=120$ GeV, $\mu=150$ GeV,
    $M_1=60.247$ GeV and by inserting the values of $M_2$, $\mu$ and
    $M_1$ into Eq.~\eqref{eq:mN}, the neutralino masses are obtained by
$$
m_{\widetilde{\chi}_{1}^{0}}=52.26~\text{GeV},
m_{\widetilde{\chi}_{2}^{0}}=90.05~\text{GeV},
m_{\widetilde{\chi}_{3}^{0}}=165.56~\text{GeV},
m_{\widetilde{\chi}_{4}^{0}}=203.49~\text{GeV}.
$$
    \item Finally, In mixture case we take $M_2= \mu=135$ GeV so obtained as
    $M_1=67.78$ GeV and also by inserting the values of $M_2$, $\mu$
    and $M_1$ into Eq.~\eqref{eq:mN}, the neutralino masses are obtained by
$$
m_{\widetilde{\chi}_{1}^{0}}=55.95~\text{GeV},
m_{\widetilde{\chi}_{2}^{0}}=95.22~\text{GeV},
m_{\widetilde{\chi}_{3}^{0}}=150.79~\text{GeV},
m_{\widetilde{\chi}_{4}^{0}}=202.40~\text{GeV}.
$$
\end{itemize}
In the numerical calculations, we use the MSTW2008 parton distribution
functions \cite{MSTW} for the quark distribution inside the proton and set
the factorization scale to the average final state mass. For each scenario given above, we have
numerically evaluated the hadronic Born cross-sections $\sigma_0$ of the process $pp\to \widetilde {\chi}_{i}^{0}\widetilde{\chi}_{j}^{0}$
(for only $u,d$ quarks and $i,j=1,2$), the EW logarithmic
contributions $\Delta\sigma$ to this process and the relative corrections $\delta$, as a function of the center-of-mass
energy from Fig.~\ref{Fig2} to Fig.~\ref{Fig4},
the $M_2$-$\mu$ mass parameters from Fig.~\ref{Fig5} to Fig.~\ref{Fig7}
and the squark mass from Fig.~\ref{Fig8} to Fig.~\ref{Fig10}, and differential cross-section as a function of the neutralino pair transverse momentum $k_T$ from Fig.~\ref{Fig11} to Fig.~\ref{Fig13}. In these figures, we use the following abbreviations:
GL, gaugino-like; HL, higgsino-like; MC, mixture-case.

In~\cref{Fig2,Fig3,Fig4}, we present the dependence of the Born level cross-sections, the EW corrections  and the relative corrections on the center-of-mass energy. These figures indicate that
both Born level cross-sections and EW corrections increase slowly and smoothly with increasing the center-of-mass
energy from 7 TeV to 14 TeV for each scenario. Furthermore, the relative corrections increase by about 2 factor as the increment of the center-of-mass energy from 7 TeV to 14 TeV. It implies that EW
contributions to the amplitudes fairly depend on the center-of-mass
energy. As shown in~\cref{Fig2}, the cross-section of the process $pp \to \widetilde
{\chi}_{1}^{0}\widetilde{\chi}_{1}^{0}$ in the higgsino-like scenario is larger than the mixing
scenario and the gaugino-like scenario in magnitude as about 42 and 65 percent, respectively. At center-of-mass energy 7 TeV (14 TeV), the EW corrections to this process increase the Born
cross-section by around 2.4$\%$ (4$\%$) in the higgsino-like scenario,
 0.6$\%$ (1.2$\%$) in the gaugino-like scenario,
0.7$\%$ (1.3$\%$) in the mixture-case scenario. Furthermore, it can be seen from
Fig.~\ref{Fig3} that the cross-section of the process $pp \to \widetilde
{\chi}_{1}^{0}\widetilde{\chi}_{2}^{0}$ in the gaugino-like scenario is larger than the mixing
scenario and the higgsino-like scenario in magnitude as about 12 and 47 percent, respectively. The EW corrections to this process increase the Born cross-section by around 2$\%$, 0.9$\%$ and 1.2$\%$ (3.5$\%$, 1.7$\%$ and 2.1$\%$) in the higgsino-like, the gaugino-like and the mixture-case scenario at center-of-mass energy 7 TeV (14 TeV), respectively. Finally, in Fig.~\ref{Fig4}, the cross-section of the process $pp \to \widetilde{\chi}_{2}^{0}\widetilde{\chi}_{2}^{0}$ in the gaugino-like scenario is larger than the mixture-case scenario and the higgsino-like scenario in magnitude as around 80 percent and 5 times, respectively. For center-of-mass energy 7 TeV (14 TeV), the EW corrections to this process increase the Born cross-section by around 5.2$\%$ (9.4$\%$) in the higgsino-like scenario, 18$\%$ (32$\%$) in the gaugino-like scenario, 15$\%$ (26$\%$) in the mixture-case scenario.

In~\cref{table1} we document a numerical survey over our scenarios for
LHC center-of-mass energies of 7 TeV and 14 TeV.
\newcommand{\thickhline}{\noalign{\hrule height 0.8pt}}
\begin{table}[htp]
\caption{The cross-sections (in fb) for the neutralino pair
production processes at
Born-level, the EW contributions to these processes and the relative correction for each scenario. Here the relative correction $\delta$ is $\Delta\sigma/ \sigma_{0}$ ratio as percent.}\label{table1}
\begin{ruledtabular}
\begin{tabular}{lcrrrrrrrrr}
~~&~&\multicolumn{3}{c}{~~~Higgsino-like}&\multicolumn{3}{c}{~~~Gaugino-like}&\multicolumn{3}{c}{~~~Mixture-case}\\ \cline{3-5} \cline{6-8} \cline{9-11}
$\sigma$~[fb]~~&~~$\sqrt{s}$~[TeV]&$\sigma_{0}$~~&$\Delta\sigma$~~&~$\delta$[\%]&~~~~~~~~$\sigma_{0}$ ~&$\Delta\sigma$~~&~$\delta$[\%]&~~~~~~~~$\sigma_{0}$~~&$\Delta\sigma$~~&~$\delta$[\%]\\
 \hline
\multirow{2}*{$pp\to\widetilde{\chi}_{1}^{0}\widetilde{\chi}_{1}^{0}$} &7&357.05&8.73&2.44&217.42&1.37&0.63&252.46&1.86&0.74\\
&14~~&800.63& 33.49&4.18 &478.73&5.61&1.17&564.47&7.61&1.35\\
\noalign{\smallskip}
\multirow{2}*{$pp\to\widetilde{\chi}_{1}^{0}\widetilde{\chi}_{2}^{0}$}
&7&13.13&0.26&1.96&19.61&0.18&0.91&17.27&0.20&1.17\\
&14~~&31.57& 1.09&3.46&45.88& 0.76&1.66&41.04&0.87&2.11\\
\noalign{\smallskip}
\multirow{2}*{$pp\to\widetilde{\chi}_{2}^{0}\widetilde{\chi}_{2}^{0}$}
&7&7.42&0.39&5.24&35.79&6.42&17.94&19.59&2.89&14.74\\
&14~~&19.93& 1.86&9.36&89.46& 28.69&32.07&50.19&13.14&26.18\\
\end{tabular}
\end{ruledtabular}
\end{table}
One can deduce from above analysis and this table that the
cross-section of the process $pp \to
\widetilde{\chi}_{1}^{0}\widetilde{\chi}_{1}^{0}$ in the
higgsino-like scenario is usually larger than others. Thus, one can
say that this process is the most dominant for neutralino pair
production processes. In particular, the cross-section of the process $pp \to \widetilde
{\chi}_{1}^{0}\widetilde{\chi}_{1}^{0}$ in the higgsino-like
scenario, appears in the range of 0.357 ($\Delta\sigma=$ 0.009) to 0.80 ($\Delta\sigma=$ 0.03) pb and should be
observable at LHC. Furthermore, for process $pp \to \widetilde
{\chi}_{2}^{0}\widetilde{\chi}_{2}^{0}$ in the gaugino-like
scenario, the cross-section appears in the range of 0.036 ($\Delta\sigma=$ 0.006) to 0.089 ($\Delta\sigma=$ 0.03) pb. Moreover, as one sees from~\cref{table1}, the EW corrections to processes $pp\to
\widetilde{\chi}_{2}^{0}\widetilde{\chi}_{2}^{0}$ are significant
and increase the Born cross-section by around 18$\%$ (32$\%$) in the
gaugino-like scenario and 15$\%$ (26$\%$) in the
mixture-case scenario for center-of-mass energy 7 TeV (14 TeV). One notes that the EW corrections are less
than for the other processes. These results imply that the relative corrections increase by about 2 factor
with increasing of the center-of-mass energy from 7 TeV to 14 TeV.

The neutralino/chargino masses and mixing matrices depend on the
$M_{2}$ and $\mu$ mass parameters, therefore one can be obtained
significant information from the dependence of the cross-section of
the neutralino pair production on these parameters.
Accordingly, we evaluate the Born level cross-sections, the EW corrections
and the relative corrections as functions of
$M_{2}$ and $\mu$ in the range from 100 to 1000 GeV in steps of 50
GeV for $\sqrt{s}=$ 8 TeV and $\tan \beta=$ 45 as displayed in ~\cref{Fig5,Fig6,Fig7}.
We can see from these figures that the Born cross-sections increase with
decreasing  $M_{2}$ and any value of $\mu$ for each process. In particular, cross-section reaches
maximal values in the region $M_{2}\lesssim 200$ GeV into the scan region. The maximum values of the relative correction are obtained in the region $\mu \lesssim 500$ GeV and $M_2=2\mu+50$(and +100) GeV for processes $pp \to \widetilde
{\chi}_{1}^{0}\widetilde{\chi}_{1}^{0}$ and $pp \to \widetilde
{\chi}_{1}^{0}\widetilde{\chi}_{2}^{0}$, whereas in the region  $\mu>M_{2}$ for process $pp \to \widetilde
{\chi}_{2}^{0}\widetilde{\chi}_{2}^{0}$. For example, it can reach about 9.5$\%$, 0.8$\%$ at $\mu=$ 300 GeV and $M_{2}=$ 650 GeV for $pp \to \widetilde{\chi}_{1}^{0}\widetilde{\chi}_{1}^{0}$, $\widetilde
{\chi}_{1}^{0}\widetilde{\chi}_{2}^{0}$, respectively, while 21.3$\%$ for $pp \to\widetilde{\chi}_{2}^{0}\widetilde{\chi}_{2}^{0}$ at $\mu=$ 650 GeV and $M_{2}=$ 300 GeV.
Furthermore, one can note that the EW correction for $pp \to \widetilde
{\chi}_{2}^{0}\widetilde{\chi}_{2}^{0}$ is larger than the remaining ones. From these figures we can see that the EW correction strongly depend on the $M_{2}$ and $\mu$ mass parameters.

In~\cref{Fig8,Fig9,Fig10}, we show the dependence
of the Born level cross-sections, the EW corrections
and the relative corrections on the squark mass for each scenario at $\sqrt{s}=$ 8 TeV. Here, there
appear the same dominant scenarios as in the dependence of the cross-sections on the center-of-mass energy. The EW corrections are not sensitive according to increment of the squark mass as shown from these figures.
It can be seen from~\cref{Fig8} that the EW corrections to $pp \to
\widetilde{\chi}_{1}^{0}\widetilde{\chi}_{1}^{0}$  increase the Born
cross-section by around 2.4$\%$, 0.6$\%$ and 0.7$\%$ in the higgsino-like, the gaugino-like and the mixture-case scenarios, respectively, for all values of the squark mass. As seen in~\cref{Fig9}, the EW corrections to $pp \to
\widetilde{\chi}_{1}^{0}\widetilde{\chi}_{2}^{0}$  increase the Born
cross-section by around 2$\%$, 0.9$\%$ and 1.2$\%$ in the higgsino-like, the gaugino-like and the mixture-case scenarios, respectively, for all values of the squark mass. Finally, the EW corrections to $pp \to
\widetilde{\chi}_{2}^{0}\widetilde{\chi}_{2}^{0}$  increase the Born
cross-section by around 5.2$\%$, 18$\%$ and 15$\%$ in the higgsino-like, the gaugino-like and the mixture-case scenarios, respectively, for all values of the squark mass as shown in~\cref{Fig10}. These results
imply that the EW corrections to $pp \to\widetilde{\chi}_{2}^{0}\widetilde{\chi}_{2}^{0}$ are larger than the others and the relative corrections are not affected by increasing of the squark mass from 400 GeV to 2000 GeV.

Finally, in~\cref{Fig11,Fig12,Fig13}, we display the dependence
of the differential cross-sections for the process $pp \to
\widetilde {\chi}_{i}^{0}\widetilde{\chi}_{j}^{0}$ as a function of
the neutralino pair transverse momentum $k_T$ at rapidity
$y_{i}=y_{j}=0$. It is seen from these figures that the differential
cross-sections reach a maximum value at around $k_T=450$ GeV and
then decrease with increasing $k_T$ in the range of 450 to 2500 GeV.
The differential cross-sections at Born-Level decrease in the range
between about $10^{-9}$ to $10^{-14}$ fb/GeV$^2$ and the
differential cross-sections of the processes with EW corrections
decrease in the range between about $10^{-10}$ to $10^{-15}$
fb/GeV$^2$ with the increment of $k_T$. It should be noted that the
dependence of the differential cross-section of the processes on the neutralino pair transverse momentum $k_T$
is dominated by one of the processes, $pp\to \widetilde {\chi}_{2}^{0}\widetilde{\chi}_{2}^{0}$ in the
gaugino-like scenario appears in the value 4.7 $\times10^{-9}$
fb/GeV$^2$. The relative correction for $pp\to\widetilde {\chi}_{1}^{0}\widetilde{\chi}_{1}^{0}$ decrease from 2.5$\%$ to 1.97$\%$,  0.68$\%$ to 0.67$\%$ and 0.78$\%$ to 0.76$\%$ in the higgsino-like, the gaugino-like and the mixture-case scenario as the increment of the transverse momentum from 300 to 2500 GeV, respectively. The relative correction for $pp\to\widetilde {\chi}_{1}^{0}\widetilde{\chi}_{2}^{0}$ decrease from 2.0$\%$ to 1.7$\%$,  0.94$\%$ to 0.83$\%$ and 1.2$\%$ to 1.0$\%$ in the higgsino-like, the gaugino-like and the mixture-case scenario with the increasing the transverse momentum from 300 to 2500 GeV, respectively. The relative correction for $pp\to\widetilde {\chi}_{2}^{0}\widetilde{\chi}_{2}^{0}$ decrease from 5.22$\%$ to 4.99$\%$,  17.8$\%$ to 17.1$\%$ and 14.6$\%$ to 13.7$\%$ in the higgsino-like, the gaugino-like and the mixture-case scenario as the increment of the transverse momentum from 300 to 2500 GeV, respectively. These results show that the EW corrections are sensitive to the transverse momentum.

\section{Conclusion}\label{Conc}

In this paper, we have considered EW corrections for the
neutralino pair production processes in proton-proton collisions at
the LHC. In the description, we have taken into account the
process $pp\to\widetilde\chi_{i}^{0}\widetilde\chi_{j}^{0}$ at the tree
level as a first choice, leading and SL
contributions for these processes at the one-loop level as a
second choice. These
corrections are significant for the theoretical and experimental
studies relating to the neutralino pair productions via the
proton-proton collisions at the LHC and the future colliders, since
they can be reach the few tens of percent level at the high energy.
We have given detail illustrations for the dependence of the cross-sections of the processes $pp \to \widetilde
{\chi}_{1}^{0}\widetilde{\chi}_{1}^{0}$,
$\widetilde{\chi}_{1}^{0}\widetilde{\chi}_{2}^{0}$,
$\widetilde{\chi}_{2}^{0}\widetilde{\chi}_{2}^{0}$, on the center-of-mass
energy, $M_2$-$\mu$ mass parameters and squark mass for three different
scenarios.

The numerical results show that the EW corrections significantly increase the Born cross-section in the dependence of the processes on the center of mass energy. In particular, the relative correction for $pp\to\widetilde {\chi}_{2}^{0}\widetilde{\chi}_{2}^{0}$ reaches about 30\% in the gaugino-like scenario. Moreover, we can see that the EW correction strongly depend on the $M_{2}$ and $\mu$ mass parameters. The maximum values of the relative correction are obtained in the region $\mu \lesssim 500$ GeV and $M_2=2\mu+50$ (and +100) GeV for processes $pp \to \widetilde
{\chi}_{1}^{0}\widetilde{\chi}_{1}^{0}$, $\widetilde
{\chi}_{1}^{0}\widetilde{\chi}_{2}^{0}$, and in the region  $\mu>M_{2}$ for process $pp \to \widetilde
{\chi}_{2}^{0}\widetilde{\chi}_{2}^{0}$. However, the squark mass dependence of the cross-sections for each scenario
decrease with increasing of the squark mass from 400 GeV to 1000
GeV, but the EW corrections are not affected by increasing of the squark mass. Finally, the dependence of the differential cross-sections for the process on the neutralino pair transverse momentum $k_T$ shows that the relative corrections decrease as the increment of the transverse momentum from 300 to 2500 GeV.

It should be underlined that there appear sizeable EW corrections to
the neutralino production, which significantly increase the
extracted bounds on the gaugino masses from the negative search for
these particles at the LHC. To our opinion these results imply an
interesting complementarity between the future LHC measurements, the
related neutralino pair measurements at a future Linear Collider. We
hope our results will be help for investigations and analysis the
different neutralino decay channels, gaugino and higgsino production
in the LHC and future hadron colliders.

\section*{Acknowledgments}
This work is supported by TUBITAK under grant number 2221(Turkey). One of the authors  A. I.~Ahmadov is grateful for financial support Baku State University Grant ``50+50''.

\end{document}